\documentclass{jfm}
\usepackage{graphicx,epstopdf,epsfig,amssymb,amsmath,psfrag,pstool,layouts}

\shorttitle{Reaction-infiltration instability}
\shortauthor{D. W. Rees Jones and R. F. Katz}

\title{Reaction-infiltration instability in a compacting porous medium}

\author{David W.~Rees Jones\aff{1}
  \corresp{\email{David.ReesJones@earth.ox.ac.uk}} \and Richard
  F.~Katz\aff{1}}

\affiliation{\aff{1}Department of Earth Sciences, University of
  Oxford, South Parks Road, Oxford, OX1 3AN, UK.}

\newcommand{\scale}[1]{\left[{#1}\right]}
\newcommand{\vell}{\boldsymbol{v}_l}
\newcommand{\vels}{\boldsymbol{v}_s}
\newcommand{\eq}{\textrm{eq}}

\begin{document}

\maketitle

\begin{abstract} 
Certain geological features have been interpreted as evidence of channelized magma flow in the mantle, which is a compacting porous medium. \citet{aharonov95} developed a simple model of reactive porous flow and numerically analysed its instability to channels. The instability relies on magma advection against a chemical solubility gradient and the porosity-dependent permeability of the porous host rock. We extend the previous analysis by systematically mapping out the parameter space. Crucially, we augment numerical solutions with asymptotic analysis to better understand the physical controls on the instability. We derive scalings for critical conditions of the instability and analyse the associated bifurcation structure. We also determine scalings for the wavelength and growth rate of the channel structures that emerge. We obtain quantitative theories for and a physical understanding of: first, how advection or diffusion over the reactive time scale set the horizontal length scale of channels; second, the role of viscous compaction of the host rock, which also affects the vertical extent of channelized flow. These scalings allow us to derive estimates of the dimensions of emergent channels that are consistent with the geologic record.
\end{abstract}

\section{Introduction} \label{sec:intro} Melting of mantle rock fuels volcanism at Hawaii and Iceland, as well as along the plate-tectonic boundaries where oceanic plates spread apart. Typically this melt is understood to come from mantle decompression: as the solid rock slowly upwells, it experiences decreasing pressure, which lowers its solidus temperature and drives quasi-isentropic melting \cite[]{ramberg72,asimow97}. The magma produced in this way segregates from its source and rises buoyantly through the interconnected pores of the polycrystalline mantle \cite[]{mckenzie84}. The equilibrium chemistry of magma is a function of pressure; rising magma, produced in equilibrium with the mantle, becomes undersaturated in a component of the mantle as it ascends \citep{OHara65, Stolper80, elthon84}. The magma reacts with adjacent solid mantle grains and the result is a net increase in liquid mass \citep{kelemen90}. This reactive melting (or, equivalently, reactive dissolution) augments decompression melting. The corrosivity of vertically segregating melt is thought to promote localisation into high-flux magmatic channels \cite[]{quick1982, kelemen92, kelemen95}; these probably correspond to zones observed in exhumed mantle rock where all soluble minerals have been replaced with olivine \citep{kelemen00, braun02}. Such channelised transport has important consequences for magma chemistry \cite[]{spiegelman03a} and, in particular, may explain the observed chemical disequilibrium between erupted lavas and the shallowest mantle \citep{kelemen95, braun02}. Laboratory experiments at high temperature and pressure confirm that magma--mantle interactions can lead to a channelisation instability \citep{pec15,pec17}. Here we analyse a simplified model of this system to better understand the character of the instability.

The association of reactive flow with channelisation was established by early theoretical work that considered a corrosive, aqueous fluid propagating through a soluble porous medium \cite[][and refs.~therein]{Hoefner88,ortoleva94}. A general feature of porous media is that permeability increases with porosity. If an increase of fluid flux enhances the dissolution of the solid matrix, increasing the porosity, then a positive feedback ensues. This drives a channelisation instability, either in the presence or absence of a propagating reaction front \cite[]{szymczak2012,Szymczak13,Szymczak14}. \cite{aharonov95} adapted the previous theory to model reactive magmatic segregation.  In their adaptation, two key differences from earlier work arise.  The first is that reaction is not limited to a moving front \cite[as in, for example,][]{hinch90}, but rather occurs pervasively within the domain. The second is that mantle rocks are ductile and undergo creeping flow in response to stress.  This includes isotropic compaction, whereby grains squeeze together and the interstitial melt is expelled (or vice versa). Equations governing the mechanics of partially molten rock were established by \cite{mckenzie84}. We will see that the compaction of the solid phase plays a crucial role in modifying and even stabilizing the instability, and so this is a key aspect of our study.

\cite{aharonov95} obtained numerical results showing the systematic dependence on reaction rate (Damk\"{o}hler number) and diffusion rate (P\'eclet number), but did not consider the co-variation of these parameters. They obtained numerical results indicative of the effect of compaction when the stiffness parameter, defined in our \S\ref{sec:eqs_simplified}, is $O(1)$. However, they did not present scalings when the stiffness parameter is much smaller than 1, which is an interesting and geologically relevant regime. \cite{spiegelman01} performed two-dimensional numerical calculations of the instability and used a similar analysis to \cite{aharonov95} to interpret the results. \cite{hewitt10} considered the reaction-infiltration instability in the context of thermochemical modelling of mantle melting. The problem was again considered by \cite{hesse11}, but their focus was mostly on an instability to compaction--dissolution waves, which were first studied by \cite{aharonov95}. While interesting theoretically, there is no geological evidence for these waves. \cite{Schiemenz2011} performed high-order numerical calculations of channelized flow in the presence of sustained perturbations at the bottom of the domain.

In the present paper, we describe the physical problem and its mathematical expression (\S\ref{sec:problem}), perform a linear stability analysis and give numerical solutions (\S\ref{sec:LSA}) and, by asymptotic analysis, elucidate the control of physical processes (\S\ref{sec:asymptotics}). The asymptotics provide scalings that are difficult to obtain numerically. They hence allow us to explore a broader parameter space, crucially including the regime in which compaction is significant. Finally, we discuss the geological implications of our analysis (\S\ref{sec:geological_discussion}).

\section{Governing equations} \label{sec:problem}

\subsection{Dimensional equations}
\begin{figure}
  \begin{center}
    \includegraphics[width=0.8\linewidth]{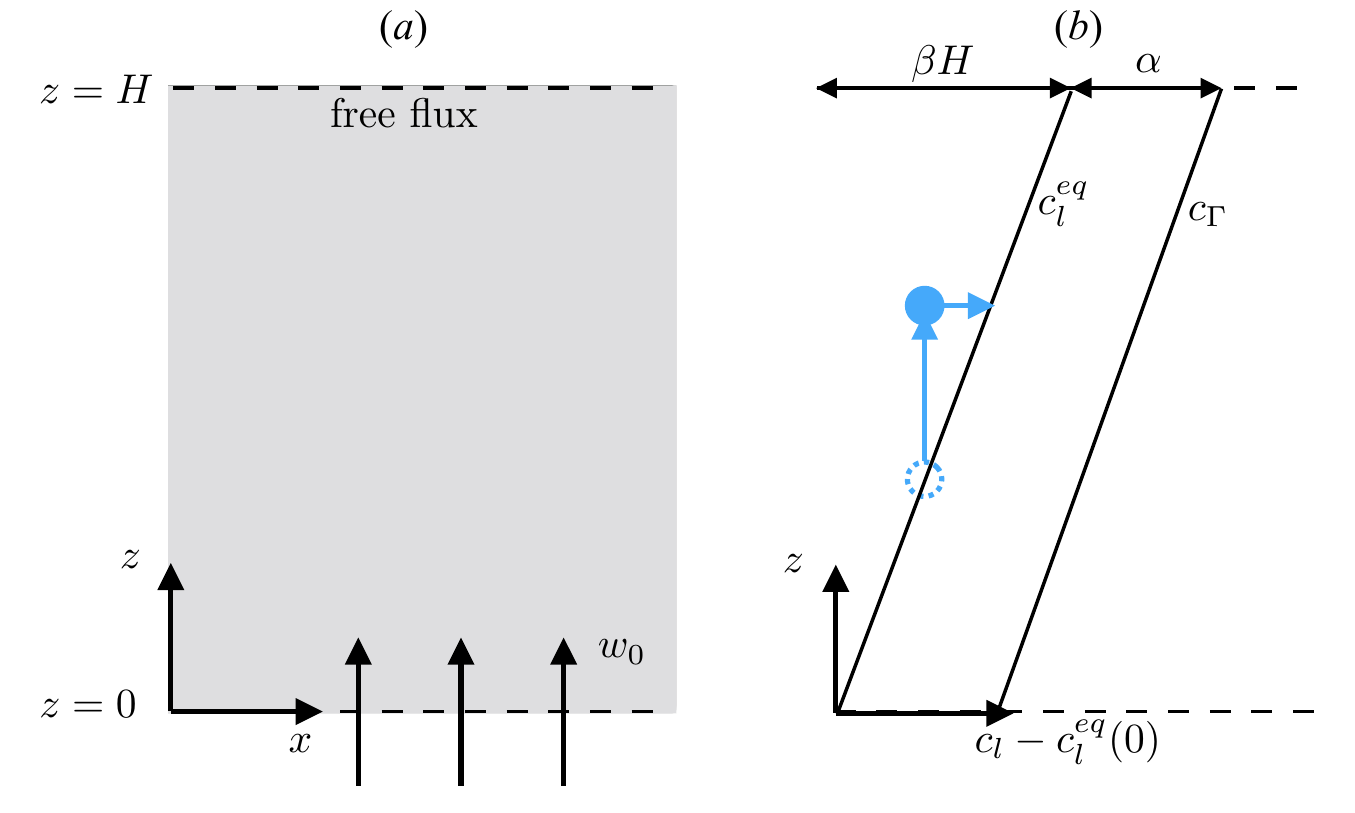}
    \caption{Diagram of the problem. (\textit{a}) shows a region of partially molten rock of depth $H$ with a flux of liquid phase from beneath and free-flux boundary condition above. (\textit{b}) shows the gradient of the equilibrium composition of the liquid phase $c^{\eq}_l $. When a parcel of liquid (dashed blue circle) is raised (full blue circle), it has a concentration below the equilibrium, leading to reactive melting along the horizontal blue arrow. The composition of reactively produced melts $c_\Gamma$ is greater than equilibrium.}
    \label{fig:diagram}
  \end{center}
\end{figure}

Figure~\ref{fig:diagram} shows a schematic diagram of the domain: a region of partially molten rock of height $H$ in the $z$ direction, composed of a solid phase ($s$, matrix, mantle rock) and a liquid phase ($l$, magma). 

We account for conservation in both phases. 
Mass conservation in the solid and liquid is given by, respectively,
\begin{subequations} 
\begin{align}
    \frac{\partial (1 -\phi)}{\partial t} + \nabla \cdot ((1-\phi) \vels)  &= -\Gamma, \label{eq:mass_s} \\
    \frac{\partial \phi}{\partial t} + \nabla \cdot (\phi \vell  ) &= \Gamma, \label{eq:mass_l}
  \end{align}
\end{subequations} 
where $t$ is time, $\phi$ is the volume fraction of liquid phase (termed porosity), $(1-\phi)$ is the fraction of solid phase, $\vell$ is the liquid velocity, $\vels$ is the solid velocity, and $\Gamma$ is the volumetric melting rate (the rate at which volume is transferred from solid to liquid phase). 

We use conservation of momentum to determine the solid and liquid velocities \citep{mckenzie84}. In general, the solid phase (mantle) can deform viscously by both deviatoric shear and isotropic compaction.  The latter is related to the pressure difference between the liquid and solid phases. We neglect deviatoric stresses on the solid phase and consider only the isotropic part of the stress and strain-rate tensors. The compaction rate $ \nabla \cdot \vels $ is related to the compaction pressure $\mathcal{P}$ according to the linear constitutive law 
 \begin{equation} \label{eq:def-zeta}
 \nabla \cdot \vels = \mathcal{P} / \zeta,
\end{equation} 
where $\zeta$ is an effective compaction or bulk viscosity. The solid matrix behaves like a rigid porous matrix when the bulk viscosity is sufficiently large (an idea we will relate to a non-dimensional matrix stiffness later). $\zeta$ can be estimated using micromechanical models of partially molten rocks, and may depend on the porosity \citep{sleep88}. The most recent calculations show that the bulk viscosity depends only weakly on porosity \citep{rudge17agu}. Therefore, we make the simplifying assumption that $\zeta$ is constant. 
We discuss this issue further in appendix~\ref{app:hewitt}.

Fluid flow is given by Darcy's law:
 \begin{equation} \label{eq:darcy_l}
  \phi \left( \vell - \vels \right) =K \left[ (1-\phi)\Delta \rho g  \hat{ \boldsymbol  z} -  \nabla \mathcal{P}  \right].
\end{equation} 
A Darcy flux $\phi \left( \vell - \vels \right) $ is driven by gravity $g  \hat{ \boldsymbol  z}$ associated with the density difference between the phases $\Delta \rho $ and by compaction pressure gradients. 
Crucially, the mobility $K$ ($\equiv$ permeability divided by liquid viscosity) of the liquid depends on the porosity:
 \begin{equation}
  K = K_0 (\phi/\phi_0)^n,
\end{equation} 
where $K_0$ is a reference mobility at a reference porosity $\phi_0$ (equal to the porosity at the base of the column $z=0$) and $n$ is a constant (we take $n=3$ in our numerical calculations). It is thought that \mbox{$2\leq n \leq 3$} for the geological systems of interest \cite[]{vonbargen86,miller14,rudge18}.

Finally, we must determine the melting or reaction rate $\Gamma$. The focus of this paper is the mechanics of the instability, so we adopt a fairly simple treatment of its chemistry, largely following \citet{aharonov95}. The reaction associated with the reaction-infiltration instability is one of chemical dissolution.  At a simple level, this can be described as follows.  As magma rises its pressure decreases and it becomes undersaturated in silica.  This, in turn, drives a reaction in which pyroxene is dissolved from the solid while olivine is precipitated \citep[\textit{cf}. figure~8 in][]{longhi02}.  Schematically, the dissolution reaction can be written:
 \begin{equation}
\mathrm{Magma}_1(l) +  \mathrm{Pyroxene}(s)  \rightarrow \mathrm{Magma}_2(l) +  \mathrm{Olivine}(s),
\end{equation}
where ($l$) denotes a component in the liquid phase and ($s$) a component in the solid phase, and we use subscript $(1,2)$ to indicate magmas of slightly different composition. Crucially, this reaction involves a net transfer of mass from solid to liquid \citep{kelemen90} and hence it is typically called a melting reaction. Because the reaction replaces pyroxene with olivine, geological observations of tabular dunite bodies in exhumed mantle rock are interpreted as evidence for the reaction-infiltration instability (dunites are mantle rocks, residual after partial melting, that are nearly pure olivine) \cite[]{kelemen92}.

We now formulate the reactive chemistry in terms of the simplest possible mathematics. We assume that $\Gamma$ is proportional to the undersaturation of a soluble component in the melt. The concentration of this component in the melt is denoted $c_l$; the equilibrium concentration is denoted $c^{\eq}_l$. Hence the melting rate is written
 \begin{equation}
  \label{eq:melting-rate-kinetic}
  \Gamma = -R\left(c_l-c^{\eq}_l\right),
\end{equation} 
where $R$ is a kinetic coefficient with units 1/time. We assume that $R$ is a constant, independent of the concentration of the soluble component in the solid phase.  This is valid for the purposes of studying the onset of instability if the soluble component is abundant and homogeneously distributed, both reasonable assumptions \cite[]{liang10}.

In this formulation, the chemical reaction rate depends on the composition of the liquid phase $c_l$. Chemical species conservation in the liquid phase is given by
 \begin{equation}
  \frac{\partial}{\partial t} \left( \phi c_l \right)  + \nabla \cdot  \left( \phi \vell c_l \right) =  \nabla \cdot \left(\phi D \nabla c_l \right) +  \Gamma  c_\Gamma, 
\end{equation} 
where the effective diffusivity of chemical species is $\phi D$ (diffusivity in the liquid phase is written $D$; diffusion through the solid phase is negligible) and $c_\Gamma$ is the concentration of reactively-produced melts. We then expand out the partial derivatives and simplify using equation~\eqref{eq:mass_l} to obtain
 \begin{equation} \label{eq:cl}
  \phi \frac{\partial c_l}{\partial t} + \phi \vell   \cdot \nabla  c_l =    \nabla \cdot \left(\phi D \nabla c_l \right) + (c_\Gamma -c_l) \Gamma. 
\end{equation} 

To close the system, we suppose that the equilibrium concentration has a constant gradient $\beta\hat{\boldsymbol{{z}}}$, as shown in figure~\ref{fig:diagram}. If we define (without loss of generality) the equilibrium concentration at the base of the region ($z=0$) to be zero, then \[c^{\eq}_l = \beta z.\] We suppose further that the concentration $c_\Gamma$ of the reactively produced melts is offset from the equilibrium concentration by $\alpha$, a positive constant, so \[c_\Gamma = \beta z + \alpha.\]  A scaling argument clarifies the meaning of the compositional parameters: for a fast reaction ($R\to\infty$) and hence for a liquid that is close to equilibrium, a vertical liquid flux $f_0$ would cause reactive melting at a characteristic rate $\Gamma_0 \sim f_0\beta/\alpha$, so $\beta/\alpha$ is the rate of reactive melting per unit of liquid flux.  Our formulation of $c_\Gamma$ is slightly different to that of \citet{aharonov95}, who take $c_\Gamma =1$. Their resulting, simplified equations are equivalent to ours when $\alpha = 1$ (following the non-dimensionalization in our \S\ref{sec:eqs_simplified}). 

At this point, we remark briefly on two simplifications inherent in the approach described above. First, we assume that the equilibrium chemistry of the liquid phase is a function of depth. A fuller treatment might consider the chemistry of the liquid as a function of pressure \cite[]{longhi02}. However, to an excellent approximation, the liquid pressure is equal to the lithostatic pressure $\rho_s g (H-z)$, in which case pressure and depth are linearly related. Indeed, the dimensionless error in making this approximation is \mbox{$O(\mathcal{S} \Delta \rho / \rho_s)$}, where $\mathcal{S}$ is the matrix stiffness parameter introduced below. Thus we neglect the difference relative to lithostatic pressure consistent with a Boussinesq approximation $\Delta \rho / \rho_s \ll 1$, where $\rho_s$ is the density of the solid phase.

Second, we use a very simple treatment of melting that neglects, for example, latent heat and temperature variations. \citet{hewitt10} developed a consistent thermodynamic model of melting and showed that latent heat may suppress instability because it reduces the melting rate. Such an effect can be represented within our simpler model by reducing the melting-rate factor $\beta/\alpha$ (see further discussion in appendix~\ref{app:hewitt}).

\subsection{Simplified, non-dimensional equations} \label{sec:eqs_simplified}
The governing equations (\ref{eq:mass_s}, \ref{eq:mass_l}, \ref{eq:darcy_l}, \ref{eq:cl}) can be non-dimensionalized according to the characteristic scales
 \begin{align}
  \label{eq:scales}
  \scale{x,z}=H, \qquad \scale{\phi} &= \phi_0,\nonumber\\
  \scale{\vell} = w_0 = K_0\Delta\rho g/\phi_0, \qquad \scale{\vels} &=
       \phi_0w_0,  \qquad \scale{t}=\alpha/\left(w_0\beta\right), \\
  \scale{\mathcal{P}}=\zeta\phi_0w_0\beta/\alpha, \qquad \scale{c_l} &= \beta H, \qquad \scale{\Gamma} =
  \phi_0w_0\beta/\alpha.\nonumber
\end{align} 
The dimensionless parameters of the system are as follows. First, $\mathcal{M} = \beta H / \alpha \ll 1$, which is the change in solubility across the domain height and characterises the reactivity of the system. Second, stiffness $\mathcal{S} = \mathcal{M} \delta^2 / H^2$, which characterises the rigidity of the medium, where $\delta = \sqrt{K_0\zeta}$ is the dimensional compaction length, an emergent lengthscale \citep[e.g.][]{spiegelman93a}. Third, $\mathrm{Da} = \alpha R H / (\phi_0 w_0) \gg 1$, the Damk\"{o}hler number, which characterises the importance of reaction relative to advection. Fourth, $\mathrm{Pe} =  w_0 H/D \gg 1$ is the P\'eclet number, which characterises the importance of advection relative to diffusion.

Then the equations can be simplified by taking the limit of small porosity $\phi_0 \ll \mathcal{M} \ll 1$ and considering only horizontal diffusion (because we expect channelized features with a short horizontal wavelength compared to their vertical structure). We also assume that the reaction rate is fast, so we neglect terms of $O(\mathcal{M}/\textrm{Da})\ll 1$. 
We also expand out the divergence term in equation~\eqref{eq:mass_s} using equation~\eqref{eq:def-zeta}.
Thus the four governing equations (\ref{eq:mass_s}, \ref{eq:mass_l}, \ref{eq:darcy_l}, \ref{eq:cl})  become
\begin{subequations}
  \label{eq:governing-non-dimensional}
\begin{align}
    \frac{\partial \phi}{\partial t} &= \mathcal{P} + \chi, \\
    \mathcal{M}  \frac{\partial \phi}{\partial t} + \nabla \cdot (\phi
    \vell  ) &=\mathcal{M} \chi, \\
        \label{eq:governing-non-dimensional-darcy}
    \phi \vell &=K \left[ \hat{ \boldsymbol z} -
       \mathcal{S} \nabla \mathcal{P} \right], \\
    \label{eq:governing-non-dimensional-concentration}
    \phi \vell   \cdot \left[\frac{ \nabla \chi
    }{\mathrm{Da}} - \hat{ \boldsymbol  z} \right] 
    &=    \frac{1}{\mathrm{Da}\mathrm{Pe}}  \frac{\partial  }
      {\partial x}\left(\phi \frac{\partial  \chi}{\partial x}\right)- \chi,
  \end{align} 
\end{subequations}
where, from this point forward, we use the same symbol to denote the dimensionless version of a variable.
The dimensionless mobility is $K=\phi^n$ and we have introduced a scaled undersaturation $\chi$ of the chemical composition of the liquid phase
 \begin{equation}
  \chi = \textrm{Da}(z-c_l). 
\end{equation} 
The dimensionless reactive melting rate is equal to the scaled undersaturation $\chi$. 

A set of appropriate boundary conditions is:
\begin{subequations}
\begin{align}
    &\phi =1, \quad  \chi =1, \quad \frac{\partial \mathcal{P}}{\partial z} = 0, \qquad (z=0),  \label{eq:nd-bcs-z0} \\
 & \frac{\partial \mathcal{P}}{\partial z} = 0,   \qquad (z=1).\label{eq:nd-bcs-z1}   \end{align} 
\end{subequations}
The boundary conditions at $z=0$ combine with equation~\eqref{eq:governing-non-dimensional-darcy} to give a incoming vertical liquid velocity $w=1$. At the upper boundary there is no driving compaction pressure gradient (a `free-flux' condition).

\section{Linear stability analysis} \label{sec:LSA}

We expand the variables as the sum of a $z$-dependent, $O(1)$ term, and a $(x,z,t)$-dependent perturbation,
\begin{subequations}
\begin{align}
    \phi &= \phi_0(z) +  \phi_1(x,z,t), \\
    \mathcal{P} &= \mathcal{P}_0(z) +  \mathcal{P}_1(x,z,t), \\
    \chi &= \chi_0(z) +  \chi_1(x,z,t), \\
    \vell &= w_0(z)\hat{\boldsymbol{z}}
      +  \boldsymbol{v}_1 (x,z,t).
  \end{align}
\end{subequations}
The perturbations are much smaller than the leading-order terms and hence we linearise the governing equations by discarding terms containing products of perturbations.

\subsection{The base state}
The leading-order flow is purely vertical. The conservation equations at this order are
\begin{subequations}
  \label{eq:zeroth_order}
\begin{align}
    0 &= \mathcal{P}_0 + \chi_0, \\
    \frac{d}{dz}  (\phi_0  w_0) &=\mathcal{M} \chi_0, \\
        \phi_0 w_0 &=K_0\left[ 1 - \mathcal{S}
      \frac{ d \mathcal{P}_0}{dz }\right], \label{eq:darcy_0} \\
    \phi_0  w_0  \left[\frac{1}{\mathrm{Da}}\frac{  d \chi_0  }{dz } -
    1\right] &=  -  \chi_0, \label{conc_0} 
  \end{align}
\end{subequations}
where $K_0=\phi_0^n$.
In the limit of large $\mathrm{Da}$, an exact solution is $\mathcal{P}_0=-\mathcal{\chi}_0$, where $\mathcal{\chi}_0=\exp(\mathcal{M}z)$. The prefactor is unity to satisfy equation~\eqref{eq:nd-bcs-z0}. We can then rearrange \eqref{eq:zeroth_order} for $\phi_0$ and $w_0$. Since $\mathcal{M} \ll 1 $, $\exp(\mathcal{M}z) \approx 1 $, 
and so we work in terms of a uniform base state,
 \begin{equation}
  \label{eq:1}
  -\mathcal{P}_0 = \chi_0 = \phi_0 = w_0 = 1.
\end{equation} 
The uniformity of the base state significantly simplifies the subsequent analysis.  

\subsection{Perturbation equations}
The equations governing the perturbations can be written
\begin{subequations}
\begin{align}
    \frac{\partial \phi_1 }{\partial t} &= \mathcal{P}_1 + \chi_1, \label{eq:solid_1} \\
    \mathcal{M}  \frac{\partial  \phi_1 }{\partial t}+ \phi_0 \nabla
    \cdot \boldsymbol v_1 + w_0 \frac{\partial \phi_1}{\partial z}   &=\mathcal{M} \chi_1, \label{eq:liquid_1} \\
     \phi_0 \boldsymbol v_1  &=-\mathcal{S} K_0 \nabla \mathcal{P}_1 +   
  (n-1) w_0 \phi_1 \hat{ \boldsymbol  z}, 
                    \label{eq:darcy_1} \\
    \left( \phi_0 w_1 + \phi_1 w_0 \right)
    \left[\frac{1}{\mathrm{Da}}\frac{
    d \chi_0  }{dz } - 1\right] +  \frac{\phi_0  w_0}{\mathrm{Da}}
    \frac{  \partial \chi_1  }{\partial z }   
      &=    \frac{\phi_0}{\mathrm{Da}\mathrm{Pe}}  
        \frac{\partial^2  \chi_1 }{\partial x^2}- \chi_1. \label{eq:reaction_1_unmod} 
          \end{align}
\end{subequations}
The third of these expressions was obtained using the exact base state relation \eqref{eq:darcy_0} and the fact that $K_0 = \phi_0^n$ and hence that $K_0' = n K_0 /\phi_0$. 

We eliminate $\chi_1$ using \eqref{eq:solid_1} and $\boldsymbol v_1$ using \eqref{eq:darcy_1}. We also use \eqref{conc_0} to simplify the expressions and obtain
\begin{subequations}
\begin{align}
    -\mathcal{S} K_0 \nabla^2 \mathcal{P}_1 +   
    n w_0 \frac{\partial \phi_1 }{\partial z}  
    &= -\mathcal{M} \mathcal{P}_1,  \label{eq:mass_1} \\
    \left(-\mathcal{S} K_0 \frac{\partial
    \mathcal{P}_1}{\partial z} + n w_0 \phi_1 \right)
    \left[\frac{ - \chi_0 }{\phi_0 w_0 }\right] 
    &=  -\left[\frac{\phi_0 w_0}{\mathrm{Da}}\frac{ \partial
      }{\partial z } -\frac{\phi_0}{\mathrm{Da}\mathrm{Pe}}
      \frac{\partial^2 }{\partial x^2}+
      1\right] \left(\frac{\partial \phi_1}{\partial t} 
      -\mathcal{P}_1\right). \label{eq:reaction_1}
  \end{align}  
\end{subequations}
We now substitute in the constant base state expressions, self-consistently neglect the $O(\mathcal{M})$ term, and cross differentiate to eliminate $\phi_1$
 \begin{equation}
  \left[ \frac{1}{\mathrm{Da}}\partial_{tz} -
    \frac{1 }{\mathrm{Da}\mathrm{Pe}}  \partial_{txx} + \partial_t  -  
    n   \right]   \nabla^2 \mathcal{P}_1      = \frac{n}{\mathcal{S}} 
  \left[ \left( \frac{1}{\mathrm{Da}} -\mathcal{S} \right)
    \partial_z
    -  \frac{1 }{\mathrm{Da}\mathrm{Pe}}  \partial_{xx} + 1\right]  
   \partial_z \mathcal{P}_1.
\end{equation}
For brevity in this equation, subscripts are used denote partial derivatives. 

We seek normal-mode solutions $\mathcal{P}_1 \propto \exp(\sigma t + ikx +mz)$ of this linear equation, where $\sigma$ is the growth rate and $k$ is a horizontal wavenumber. Thus we obtain the characteristic polynomial (dispersion relationship)
 \begin{equation} \label{eq:dispersion3} \frac{\sigma}{\mathrm{Da}} m^3
  + \left( \sigma \mathcal{K} - \frac{n}{\mathrm{Da} \mathcal{S}}
  \right)m^2 - \left( \frac{n\mathcal{K}}{\mathcal{S}} +
    \frac{\sigma}{\mathrm{Da}} k^2 \right) m + \left(n-\sigma
    \mathcal{K} \right)k^2 = 0,
\end{equation} 
where $\mathcal{K}=1+k^2/\mathrm{Da}\mathrm{Pe}$.  Equation~\eqref{eq:dispersion3} has three roots $m_j$ $(j=1,2,3)$ and hence the compaction pressure perturbation will be given by
 \begin{equation}
  \label{eq:cmppres_solution}
  \mathcal{P}_1 = \sum_{j=1}^3 A_j\exp(\sigma t + ikx + m_jz).
\end{equation} 
The three unknown pre-factors $A_j$ are determined by the boundary conditions.

\subsection{Boundary conditions on the perturbation} \label{sec:bcs}
We previously eliminated $\chi_1$ and $\phi_1$ in favour of the compaction pressure $\mathcal{P}_1$. The corresponding boundary conditions on  $\mathcal{P}_1$, derived from equations~\eqref{eq:nd-bcs-z0} and \eqref{eq:solid_1} are
\begin{subequations}
  \label{eq:boundary_conditions}
\begin{align}
  \mathcal{P}_1 &= 0 \quad \mathrm{at } \, z=0, \label{eq:pert_bc_1} \\
   \frac{\partial \mathcal{P}_1}{\partial z}  &= 0 \quad \mathrm{at } \, z=0. \label{eq:pert_bc_2}
  \end{align} 
The upper boundary condition (equation~\ref{eq:nd-bcs-z1}) is
\begin{equation}
    \frac{\partial \mathcal{P}_1}{\partial z}  = 0 \quad \mathrm{at } \, z=1.  \label{eq:pert_bc_3}
  \end{equation} 
\end{subequations}
The boundary conditions can be expressed in matrix form in terms of the coefficients of the normal-mode expansion~\eqref{eq:cmppres_solution} as
 \begin{equation}
  \label{eq:bc_matrix}
  \left(\begin{array}{ccc}
      1 & 1 & 1 \\
      m_1 & m_2 & m_3 \\
      m_1\textrm{e}^{m_1} & m_2\textrm{e}^{m_2} & m_3\textrm{e}^{m_3}     
    \end{array}\right)
  \left(\begin{array}{c}
      A_1 \\ A_2 \\ A_3
    \end{array}\right) = 
  \left(\begin{array}{c}
      0 \\ 0 \\ 0
    \end{array}\right) .
\end{equation} 
A necessary (but not sufficient condition) for a non-trivial solution $A_j$ to exist is that the boundary-condition matrix $M$ has zero determinant.

\subsection{Analysis of the dispersion relationship} \label{sec:analysis1}
We analyse the characteristic polynomial~\eqref{eq:dispersion3} for the case of real growth rate $\sigma$ (that is, we look for channel modes rather than compaction-dissolution waves, as discussed in \S\ref{sec:intro}). The characteristic polynomial, a cubic, has three roots $m_j$ $(j=1,2,3)$. The character of these roots is controlled by the cubic discriminant. If the discriminant is strictly positive, the roots are distinct and real. If the discriminant is zero, the roots are real but at least one root is repeated (degenerate).  If the discriminant is strictly negative, then there is one real root ($m_1$, say), and a pair of complex conjugate roots ($m_2,m_3$). 

For the case of real and distinct roots, the columns of $M$ are linearly independent, the determinant of $M$ is non-zero, and the only solution has $A_j=0$. When the roots are real but degenerate, $\det M = 0$ but there is no set of coefficients $A_j$ that can satisfy the boundary condition at $z=1$~\eqref{eq:pert_bc_3}. Hence there are physically meaningful roots only when the cubic discriminant of~\eqref{eq:dispersion3} is strictly negative.

In this latter case, with one real root and two complex conjugate roots, $\det M$ is purely imaginary. A proof of this follows. Consider a $2 \times 2$ matrix whose columns are complex conjugate, say $Y= \left( \boldsymbol{X}, \boldsymbol{X}^* \right)$ where $\boldsymbol{X}=[X_1,X_2]^T$. Then $\det Y = X_1 X_2^* - X_2 X_1^*$, so $\det Y + \det Y^* = 0$, i.e., $\det Y$ is pure imaginary. The boundary condition matrix $M$ is $3\times 3$, but $\det M$ can be written as the sum of purely imaginary determinants of $2 \times 2$ sub-matricies, multiplied by purely real numbers; hence $\det M$ is pure imaginary. 

With $m_1$ real, and $m_2$ and $m_3=m_2^*$ complex, there are eigenvalues of $\sigma$ for which the imaginary part of $\det M$ vanishes. At these eigenvalues, $\det M = 0$ and there exists an eigenvector $A_j$ such that the boundary conditions are satisfied. We find these eigenvalues/vectors by numerically solving the coupled problem of the cubic polynomial~\eqref{eq:dispersion3} and $\det M = 0$. 

\subsection{Physical discussion of instability mechanism (part I: growth rate)} \label{sec:physical-discussion-1}

\begin{figure}
  \begin{center}
    \includegraphics[width=1.0\linewidth]{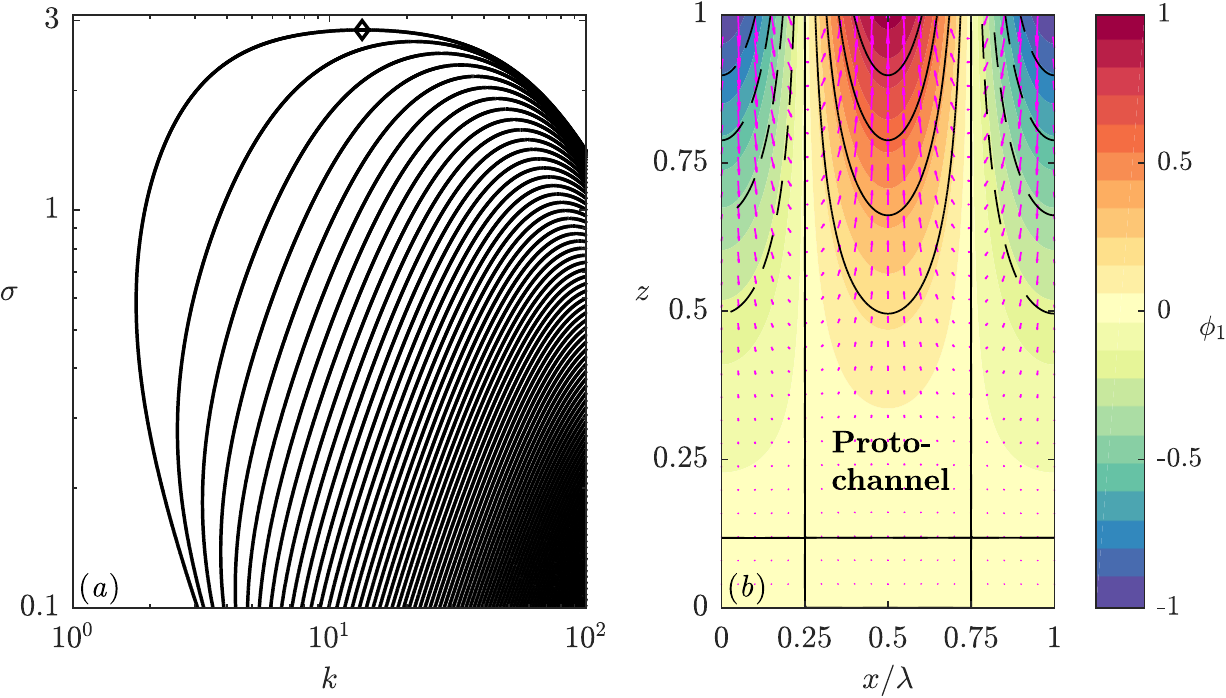}
    \caption{(\textit{a}) Example dispersion relationship calculated at $\mathrm{Da}=10^2$, $\mathrm{Pe}=10^2$, $\mathcal{S}=1$, $n=3$. (\textit{b}) Perturbation corresponding to most unstable wavenumber (indicated by a diamond symbol in panel (\textit{a})). The background colour scale shows the porosity perturbation $\phi_1$ (normalized to have a maximum value of 1). The black curves are contours of liquid undersaturation $\chi_1$, which is positively correlated with $\phi_1$ (solid = positive, dashed = negative). The magenta arrows show the perturbation liquid velocity $\boldsymbol v_1$. Note the flow into the proto-channels (regions of elevated porosity $\phi_1>0$). The compaction pressure $\mathcal{P}_1$ (not shown) is anti-correlated with $\phi_1$, consistent with flow direction from high to low pressure.}
    \label{fig:perturbation_eg}
  \end{center}
\end{figure}

Figure~\ref{fig:perturbation_eg}(\textit{a}) shows an example of the dispersion relationship $\sigma(k)$. The curves are a series of valid solutions. The solutions on the uppermost dispersion curve have the largest growth rate $\sigma$ and are monotonic in $z$.  Curves below this fundamental mode are higher order, with increasing numbers of turning points in $z$ as $\sigma$ decreases at fixed $k$. In this example, the instability is only present at $k \gtrsim 1$, which roughly translates to channels that are narrower than the domain height.  Hence we expect that the lateral wavelength is always smaller than the domain height.

We now explain the physical mechanism that gives rise to the instability.
Figure~\ref{fig:perturbation_eg}(\textit{b}) shows an example of the structure of the fastest-growing perturbation (most unstable mode). 
Regions of positive porosity perturbation $\phi_1$ (which we call proto-channels) create a positive perturbation of the vertical flux, according to equation~\eqref{eq:darcy_1}. For didactic purposes, consider the case of no compaction pressure (which is directly applicable to a rigid porous medium). Then 
 \begin{equation} \label{eq:scaling_1}
 \phi_0 w_1 + \phi_1 w_0  = n w_0 \phi_1.
\end{equation} 
Note that the positive vertical flux perturbation only occurs because the permeability increases with porosity ($n>0$); this is a crucial aspect of the instability.

Positive vertical advection against the background equilibrium concentration gradient leads to positive liquid undersaturation $\chi_1$, according to equation~\eqref{eq:reaction_1_unmod}. 
In more physical terms, the enhanced vertical flux advects corrosive liquid from below. 
Thus the equilibrium concentration gradient is the other crucial aspect of the instability, alongside the porosity-dependent permeability.
For didactic purposes, consider the case of very fast reaction ($\mathrm{Da} \gg 1$), in which the leading order balance in equation~\eqref{eq:reaction_1_unmod} gives
 \begin{equation} \label{eq:scaling_2}
\chi_1 = \phi_0 w_1 + \phi_1 w_0  = n w_0 \phi_1.
\end{equation} 
Positive liquid undersaturation in turn causes reactive melting and hence increasing porosity by equation~\eqref{eq:solid_1}, so the proto-channel emerges. Again, neglecting compaction pressure, replacing $\partial_t \to \sigma$, and substituting equation~\eqref{eq:scaling_2}, we find
 \begin{equation} \label{eq:scaling_3}
\sigma \phi_1 = n w_0 \phi_1 \quad \Rightarrow \quad  \sigma = n,
\end{equation} 
where we used $w_0=1$. Note that the maximum growth rate in Figure~\ref{fig:perturbation_eg}(\textit{a}) is about $n=3$. 
Recalling the non-dimensionalization of time in equation~\eqref{eq:scales}, we see that the timescale for channel growth is the timescale for reactive melting ($\alpha /\beta w_0$)  multiplied by the sensitivity of melt flux to porosity ($n$).

Further consideration of equation~\eqref{eq:reaction_1_unmod} reveals two stabilising mechanisms. The instability is weakened by diffusion, especially at high wavenumber, since diffusion acts to smooth out lateral gradients in the undersaturation. It is also weakened by advection of the liquid undersaturation, because the undersaturation in the proto-channel increases with height \mbox{($\partial \chi_1/\partial z >0$)}. The subsequent analysis shows that this latter mechanism is also more important at large wavenumber, so both advection and diffusion of liquid undersaturation play a role in wavelength selection (see \S\ref{sec:advection-controlled}  and \S\ref{sec:diffusion-controlled}, respectively). Indeed, figure~\ref{fig:perturbation_eg}(\textit{a}) shows that the growth rate decreases at large $k$.

Finally, we consider the effect of compaction, which is a further stabilising mechanism at both large and small wavenumbers \citep{aharonov95} (see \S\ref{sec:wavenumber}, \S\ref{sec:compaction} and appendix~\ref{app:kminmax}). 
The instability only occurs if the matrix stiffness exceeds some critical value (see \S\ref{sec:Scrit} and \S\ref{sec:Scrit_analysis}).
To leading order ($\mathcal{M} \ll 1$), if we consider equation~\eqref{eq:liquid_1} governing liquid mass conservation, then
 \begin{equation} \label{eq:scaling_4}
  \phi_0 \nabla \cdot \boldsymbol v_1 = - w_0 \frac{\partial \phi_1}{\partial z} 
 \quad  \Rightarrow  \quad K_0 \nabla^2 \mathcal{P}_1 =  
  \frac{n w_0}{\mathcal{S}} \frac{\partial \phi_1}{\partial z},  
\end{equation} 
where we substitute in equation~\eqref{eq:darcy_1} to achieve the last expression (\textit{cf}. equation~\ref{eq:mass_1}).    
Proto-channels are regions of increasing porosity perturbation \mbox{($\partial \phi_1/\partial z>0$)}.  
Thus, by liquid mass conservation, they are regions of convergence of the perturbation velocity $\boldsymbol v_1$. 
Therefore, proto-channels are regions of negative compaction pressure perturbation, which reduces the porosity perturbation, according to the equation of solid mass conservation \eqref{eq:solid_1}. 
Again, this stabilising mechanism is wavelength dependent through the Laplacian in equation~\eqref{eq:scaling_4}. Note further that the perturbation to the compaction pressure decreases with increasing matrix stiffness $\mathcal{S}$, so we recover the rigid porous medium case as $\mathcal{S}\gg1$. 
We return to the physical discussion of the instability in \S\ref{sec:physical-discussion-2} to explain the wavelength selection and the critical matrix stiffness.

\section{Asymptotic analysis of the large-$\boldsymbol{\mathrm{Da}}$ limit}  \label{sec:asymptotics}
In this section, we use asymptotic analysis to estimate the maximum growth rate $\sigma^*$ and the the wavenumber $k^*$ of the most unstable mode. The analysis allows us to understand the physical controls on the instability, particularly the wavelength selection.

The cubic dispersion relation \eqref{eq:dispersion3} has a structure that simplifies in the limit of large $\mathrm{Da}$. There is one real root of $O(\mathrm{Da})$ and a pair of complex conjugate roots. Take $m_1\sim O(\mathrm{Da})$ as ansatz and obtain:
 \begin{equation}
  m_1 \sim -  \mathcal{K} \mathrm{Da} .
\end{equation} 
Take $m_{2,3} \sim O(\mathrm{1})$ as ansatz and obtain:
 \begin{equation} \label{eq:dispersion2} \left( \sigma \mathcal{K} -
    \frac{n}{\mathrm{Da} \mathcal{S}} \right)m^2 - \left(
    \frac{n\mathcal{K}}{\mathcal{S}} + \frac{\sigma}{\mathrm{Da}} k^2
  \right) m + \left( n-\sigma \mathcal{K} \right)k^2 = 0.
\end{equation} 

The boundary condition~\eqref{eq:pert_bc_2} is accommodated by a boundary layer of thickness $O(1/\mathrm{Da})$ associated with the root $m_1$. The remaining boundary conditions~(\ref{eq:pert_bc_1} \& \ref{eq:pert_bc_3}) can be written
 \begin{equation}
  \label{eq:bc_matrix_quadratic}
  \left(\begin{array}{cc}
       1 & 1 \\
      m_2\textrm{e}^{m_2} & m_3\textrm{e}^{m_3}     
    \end{array}\right)
  \left(\begin{array}{c}
      A_2 \\ A_3
    \end{array}\right) = 
  \left(\begin{array}{c}
      0 \\ 0
    \end{array}\right) .
\end{equation} 
As before, we require the determinant of this boundary condition matrix to be zero. Noting that $m_3=m_2^*$, we find that
 \begin{equation}
  0 = \mathrm{imag} \left[ m_2 \exp(m_2) \right] . 
\end{equation} 
We write $m_2$ in terms of its real and imaginary parts $m_2=a+i b$, then
 \begin{equation} \label{eq:critical_condition_2} 0 = \tan b + b/a.
\end{equation} 
This algebraic equation has an infinite family of solutions corresponding to the multiple roots shown in figure~\ref{fig:perturbation_eg}(\textit{a}).
The perturbation compaction pressure can be written
 \begin{equation}
  \mathcal{P}_1 \propto \exp(az) \sin(bz).
\end{equation} 
Note that there is no part of the solution proportional to \mbox{$ \exp(az) \cos(bz)$} because of boundary condition~\eqref{eq:pert_bc_1}. Equation \eqref{eq:critical_condition_2} is equivalent to boundary condition~\eqref{eq:pert_bc_3}. 

\begin{figure}
  \begin{center}
    \includegraphics[width=3.4950in]{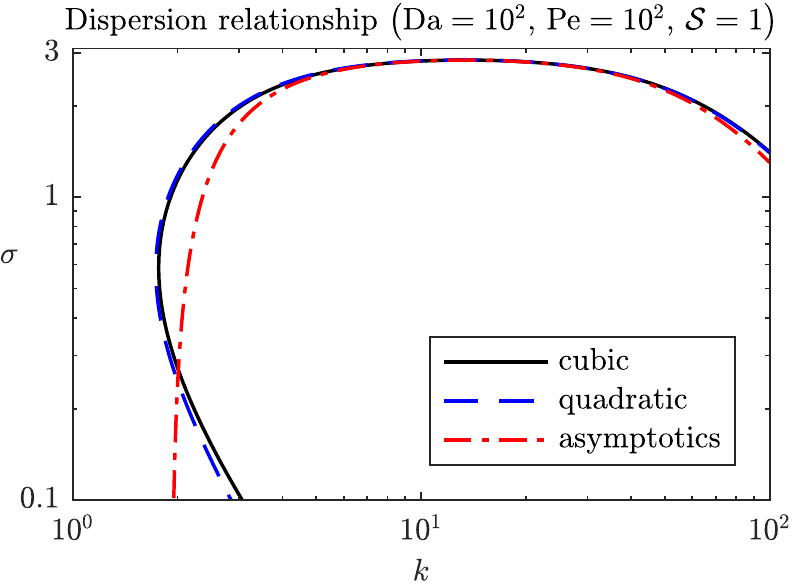}
    \caption{High $\mathrm{Da}$ dispersion relationship with scaling relationships overlaid. Solid black: full numerical calculation of cubic dispersion relationship~\eqref{eq:dispersion3}, only showing the most unstable mode. Dashed blue:   solution of equations (\ref{eq:ab_full_1}, \ref{eq:ab_full_2}) from the simplified quadratic dispersion relationship. Dot-dashed red: solution of equation~\eqref{eq:ae_full_2}. The blue curve agrees well everywhere, the red curve is only valid when $n-\sigma$ is small, consistent with the asymptotic approximations.  }
    \label{fig:dispersion_high_da}
  \end{center}
\end{figure}

The real and imaginary parts of $m_2$ can be found using a variant of the quadratic formula. 
 \begin{equation} \label{eq:quadratic_formula}
px^2+qx+r=0 \Rightarrow x= \frac{-q}{2p} \pm i \sqrt{\frac{r}{p}-\left(\frac{q}{2p} \right)^2},
\end{equation} 
where we assume that the quantity within the square root is real for the reasons discussed above (\S\ref{sec:analysis1}).
We use equation~\eqref{eq:quadratic_formula} to obtain the exact expressions
\begin{subequations} \label{eq:ab_full_1+2} 
\begin{align} \label{eq:ab_full_1}
    a &=  \frac{\left(     \frac{n\mathcal{K}}{\mathcal{S}} + 
        \frac{\sigma}{\mathrm{Da}} k^2  \right)}{{2  \left( \sigma
        \mathcal{K} - \frac{n}{\mathrm{Da} \mathcal{S}}  \right)}}, \\
    b^2 + a^2 &=\frac{\left( n-\sigma \mathcal{K} \right) k^2}{{
                \left( \sigma \mathcal{K} - \frac{n}{\mathrm{Da}
                \mathcal{S}} \right)}} .  \label{eq:ab_full_2}
  \end{align} 
\end{subequations}
It is possible to solve these algebraic equations for $\sigma$ numerically (\textit{cf}. dashed blue curve in figure~\ref{fig:dispersion_high_da}), but it is instructive to make the additional ansatz $\sigma \sim n(1-\epsilon)$, where $\epsilon \ll 1$. This allows us to approximate the behaviour near the maximum growth rate, where $\sigma \sim n$. We also assume that $(\mathrm{Da} \mathcal{S})^{-1} \ll 1$ but retain terms $O(k^2 \mathrm{Da} ^{-1})$ since the latter is important at large wavenumber. In general $\mathcal{K} \sim 1$, except on the right-hand-side of equation~\eqref{eq:ab_full_2}, where we obtain a term proportional to $\left[(1-\epsilon)^{-1} - \mathcal{K}\right] \sim \epsilon - k^2/\mathrm{Da} \mathrm{Pe}$.  We test all the results obtained using these approximations against full numerical solution of the cubic dispersion relation. Under the simplifying assumptions,
\begin{subequations} \label{eq:ae_full_1+2} 
\begin{align} \label{eq:ae_full_1}
    a & \sim      \frac{1}{2 \mathcal{S}} + \frac{ k^2 }{2 \mathrm{Da}}, \\
    \epsilon &\sim \frac{b^2 + a^2}{k^2} + 
               \frac{k^2}{\mathrm{Da} \mathrm{Pe} } .  \label{eq:ae_full_2}
  \end{align}
\end{subequations}
The terms that constitute $\epsilon$ represent a series of stabilizing mechanisms that reduce the growth rate $\sigma$, namely compaction (through the $1/2\mathcal{S}$ term in equation~\eqref{eq:ae_full_1}), advection of undersaturation (through the $k^2/2\mathrm{Da}$ term in equation~\eqref{eq:ae_full_1}), and diffusion (through the $k^2/\mathrm{Da} \mathrm{Pe}$ term in equation~\eqref{eq:ae_full_2}).
We show an example dispersion relationship at moderately high $\mathrm{Da}$ in figure~\ref{fig:dispersion_high_da}.

\subsection{Dependence on wavenumber $k$} \label{sec:wavenumber}
Starting at small $k$, $ \epsilon $ initially decreases with $k$, reaches some minimum value $ \epsilon^* $ at $k=k^*$ [corresponding to the most unstable mode with maximum growth rate $\sigma^* =n(1-\epsilon^*)$], and then increases as $k \rightarrow \infty$.

Scaling arguments make these statements more precise. When
$k \ll k^*$,
\begin{subequations}
\begin{align}
    a & \sim      \frac{1}{2 \mathcal{S}}  \label{eq:ae_smallk_1}  \\
    \epsilon &\sim \frac{b^2 +
               a^2}{k^2}. \label{eq:ae_smallk_2}
  \end{align}
\end{subequations}
It is convenient to define $\mathcal{B}(\mathcal{S}) = b^2 + a^2$, where $a = 1/2\mathcal{S}$ and $b$ satisfies equation \eqref{eq:critical_condition_2}. Then a small wavenumber `cut-off' occurs when $\epsilon = O(1)$ (which is outside the bounds of our previous assumption $\epsilon \ll 1$) when $k \sim \mathcal{B}^{1/2}$. 
We use `cut-off' to refer to the wavenumber at which the growth rate departs significantly from its maximum value, not the strict minimum wavelength, which we discuss below. 

Conversely, at large $k$, $k \gg k^*$
\begin{subequations}
\begin{align}
    a & \sim      \frac{1}{2 \mathcal{S}} + \frac{k^2}{2      \mathrm{Da}}  , \label{eq:ae_largek_1} \\
    \epsilon &\sim k^2
               \left({\frac{1}{(2\mathrm{Da})^2}}
               +\frac{1}{\mathrm{Da}\mathrm{Pe}}\right)
               .  \label{eq:ae_largek_2}
  \end{align}
\end{subequations}
If $\mathrm{Da} \gg \mathrm{Pe}$, then $ \epsilon \sim {k^2}/{\mathrm{Da} \mathrm{Pe} } $, so the large wavenumber `cut-off' occurs when $k \sim (\mathrm{Da} \mathrm{Pe})^{1/2}$. Physically, the small scale of the instability is limited by the distance a chemical component can diffuse over the reaction timescale \citep{spiegelman01}. Conversely if $\mathrm{Pe} \gg \mathrm{Da}$, then $ \epsilon \sim {k^2}/{(2 \mathrm{Da} )^2 } $, so the large wavenumber `cut-off' occurs when $k \sim 2 \mathrm{Da}$. Physically, the small scale of the instability is limited by the distance a chemical component is transported by the background liquid flow over the reaction timescale. These two limits also affect the maximum growth rate of the instability. In the next sections we consider each limit in turn.

It is also possible to determine strict minimum and maximum wavenumbers for instability, although this is more technical so we leave the details for appendix~\ref{app:kminmax}. In summary, we find
\begin{align}
&k_\mathrm{min} \sim \frac{1.5171}{ \mathcal{S}^{1/2}}  \quad (\mathcal{S} \gg 1),
\qquad k_\mathrm{min} \sim \frac{1}{ \mathcal{S}} \quad  (\mathcal{S} \ll  1), \\
&k_\mathrm{max} \sim \mathcal{S}  \mathrm{Da}  \mathrm{Pe}.
\end{align}
The dependence on matrix stiffness $\mathcal{S}$ means that compaction stabilizes the system at both large and small wavenumbers \citep{aharonov95}. 
Indeed, in a rigid medium \mbox{($\mathcal{S} \gg 1$)} there is no minimum or maximum wavenumber. 

\subsection{Advection controlled instability $\mathrm{Pe} \gg \mathrm{Da} \gg 1$} \label{sec:advection-controlled}
We first consider case of negligible diffusion. In this case, it is 
natural to introduce a change of variables:
$\tilde{k} = k \mathrm{Da}^{-1/2}$,
$\tilde{\epsilon} = \epsilon \mathrm{Da}$. Then, to leading order,
\begin{subequations}
\begin{align}
    a &\sim   \frac{1}{2\mathcal{S}} + \frac{\tilde{k}^2}{2}, \\
    \tilde{\epsilon} &\sim \frac{a^2 +
                       b^2}{\tilde{k}^2}. \label{eq:epsilon-tilde}
  \end{align}
\end{subequations}
Note that both $b$ and $a$, and hence $\tilde{\epsilon}$, are functions
of $(\tilde{k},\mathcal{S})$ alone.

\begin{figure}
  \begin{center}
    \includegraphics[width=1.0\linewidth]{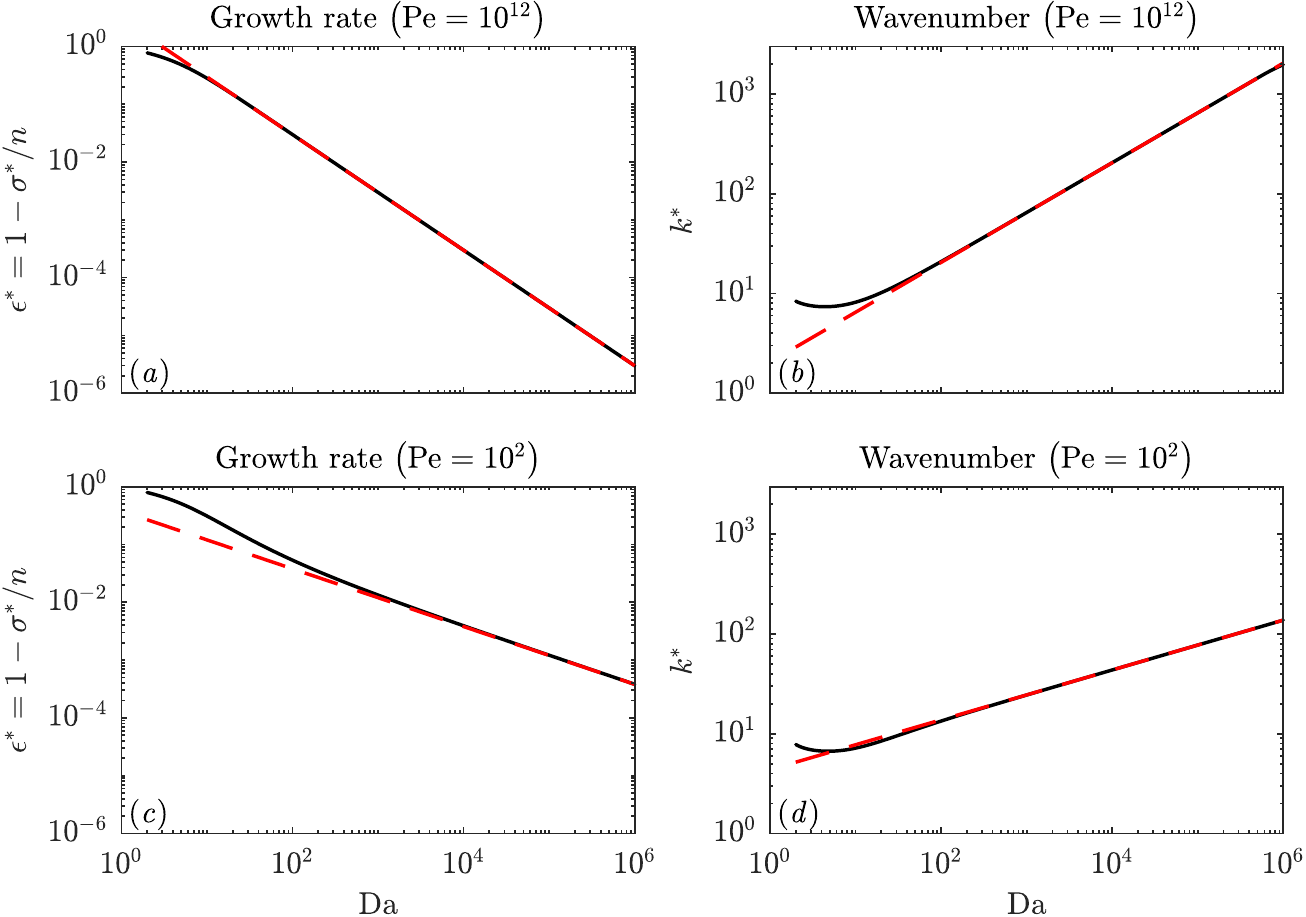}
    \caption{Numerical calculations of the full cubic dispersion relation~\eqref{eq:dispersion3} (solid black curves) compared to power-law scalings (dashed red lines) for maximum growth rate and corresponding wavenumber as a function of $\mathrm{Da}$, at fixed $\mathrm{Pe}=10^{12}$ (panels \textit{a, b}), and $\mathrm{Pe}=10^2$ (panels \textit{c, d}). For all calculations $\mathcal{S}=1$.}
    \label{fig:Da_scalings}
  \end{center}
\end{figure}

We find the maximum growth rate by differentiating equation \eqref{eq:epsilon-tilde} and seeking the (unique) turning point, which satisfies \begin{equation} \label{eq:k-tilde}
  b^2 + \left(\tilde{k}^2-1 \right) b \cos(b)\sin(b) -\tilde{k}^2 \sin^2(b) =0.
\end{equation} 
We solve numerically to obtain the solution $\tilde{k}^*=\tilde{k}^*(\mathcal{S})$. The corresponding growth rate is $\tilde{\epsilon}^*(\mathcal{S})$. In summary, the most unstable wavelength $k^* \sim \tilde{k}^* \mathrm{Da}^{1/2}$ \citep[consistent with the numerical results of][]{aharonov95}  and the corresponding growth rate $\sigma^* \sim n\left[1-\tilde{\epsilon}^*\mathrm{Da}^{-1}\right]$. These scaling results are shown in figure~\ref{fig:Da_scalings}  (panels \textit{a, b}). The dependence on compaction through matrix stiffness $(\mathcal{S})$ is shown in figure~\ref{fig:S_scalings}  (panels \textit{a, b}).
The wavenumber is controlled by advection of liquid undersaturation 
(see \S\ref{sec:physical-discussion-2}).

\subsection{Diffusion controlled instability $\mathrm{Da} \gg \mathrm{Pe}$}  \label{sec:diffusion-controlled}
The other limit occurs when diffusion is significant.  For this case, it is natural to introduce a different change of variables: $\hat{k} = k (\mathrm{Da}\mathrm{Pe})^{-1/4}$, $\hat{\epsilon} = \epsilon (\mathrm{Da} \mathrm{Pe})^{1/2} $. Then
\begin{subequations}
\begin{align}
    a &\sim   \frac{1}{2\mathcal{S}}, \\
    \hat{\epsilon} & \sim
                     \frac{\mathcal{B}+\hat{k}^4}{
                     \hat{k}^2},
  \end{align} 
\end{subequations}
where $\mathcal{B}(\mathcal{S})$ was defined previously.

This dispersion relation is simple enough to analyse by hand. The minimum of $\hat{\epsilon}^*=2\mathcal{B}^{1/2}$ and occurs when $\hat{k}^*=\mathcal{B}^{1/4}$. Thus the maximum growth rate $\sigma^*$ that occurs at wavenumber $k^*$ satisfies
\begin{subequations}
\begin{align}
    k^* &\sim (\mathrm{Pe Da}\mathcal{B} )^{1/4},\\
   \sigma^* &\sim n
                          \left[1-\frac{2
                          \sqrt\mathcal{B}}{\sqrt{\mathrm{Da
                          Pe}}}
                          \right].
  \end{align}
\end{subequations}
That $ k^* \sim \mathrm{Pe }^{1/4}$ was observed numerically by \citet{aharonov95}, although they did not obtain the dependence on $\mathrm{Da}$ or $\mathcal{S}$. 
Thus the instability grows most rapidly at some wavelength controlled by diffusion. The analysis is consistent with numerical results (figure~\ref{fig:Da_scalings}\textit{c, d}). The dependence on compaction through the function  $\mathcal{B}(\mathcal{S})$ is shown in figure~\ref{fig:S_scalings}  (panels \textit{c, d}).
Increasing matrix stiffness $\mathcal{S}$ increases the growth rate and reduces the wavenumber of the most unstable mode. 
The wavenumber is controlled by diffusion 
(see \S\ref{sec:physical-discussion-2}).

\subsection{Effect of compaction (dependence on $\mathcal{S}$)} \label{sec:compaction}
Asymptotic estimates of the dependence on $\mathcal{S}$ are obtained by analysing the roots of equation \eqref{eq:critical_condition_2}: $\tan b + b/a = 0$. The first non-trivial root $b$ of this equation occurs for $b \in (\pi/2, \pi)$. At small $a$ (large $\mathcal{S}$), the root $b \rightarrow  \pi/2^+ $. At large $a$ (small $\mathcal{S}$), the root $b \rightarrow \pi^-$.

Next we determine the maximum growth rate for large and small $\mathcal{S}$. First we consider the case of advection controlled growth ($\mathrm{Pe} \gg \mathrm{Da} \gg 1$). At large $\mathcal{S} \gg 1$, $a \sim \tilde{k}^2/2$ independent of $\mathcal{S}$. Thus $a,b,\tilde{k}^*,\tilde{\epsilon}^*$ approach some limit that is independent of $\mathcal{S}$. By solving equation~\eqref{eq:k-tilde} numerically, we find that  
\begin{subequations}
\begin{align}
    \tilde{k}^* &\rightarrow 1.898, \\
    \tilde{\epsilon}^* &\rightarrow 2.302.  \label{eq:highDa_infinitePe_highS_limit}
  \end{align}
\end{subequations}
At small $\mathcal{S} \ll 1$, we obtain the following leading order expressions
\begin{subequations}
\begin{align}
    a^* &\sim \mathcal{S}^{-1}, \\
    b^* &\sim \pi(1- \mathcal{S} ), \\
    \tilde{k}^* &\sim \mathcal{S}^{-1/2}, \\
    \tilde{\epsilon}^* &\sim
                         \mathcal{S}^{-1}.  \label{eq:highDa_infinitePe_S_scaling}
  \end{align}
\end{subequations}

\begin{figure}
  \begin{center}
    \includegraphics[width=1.0\linewidth]{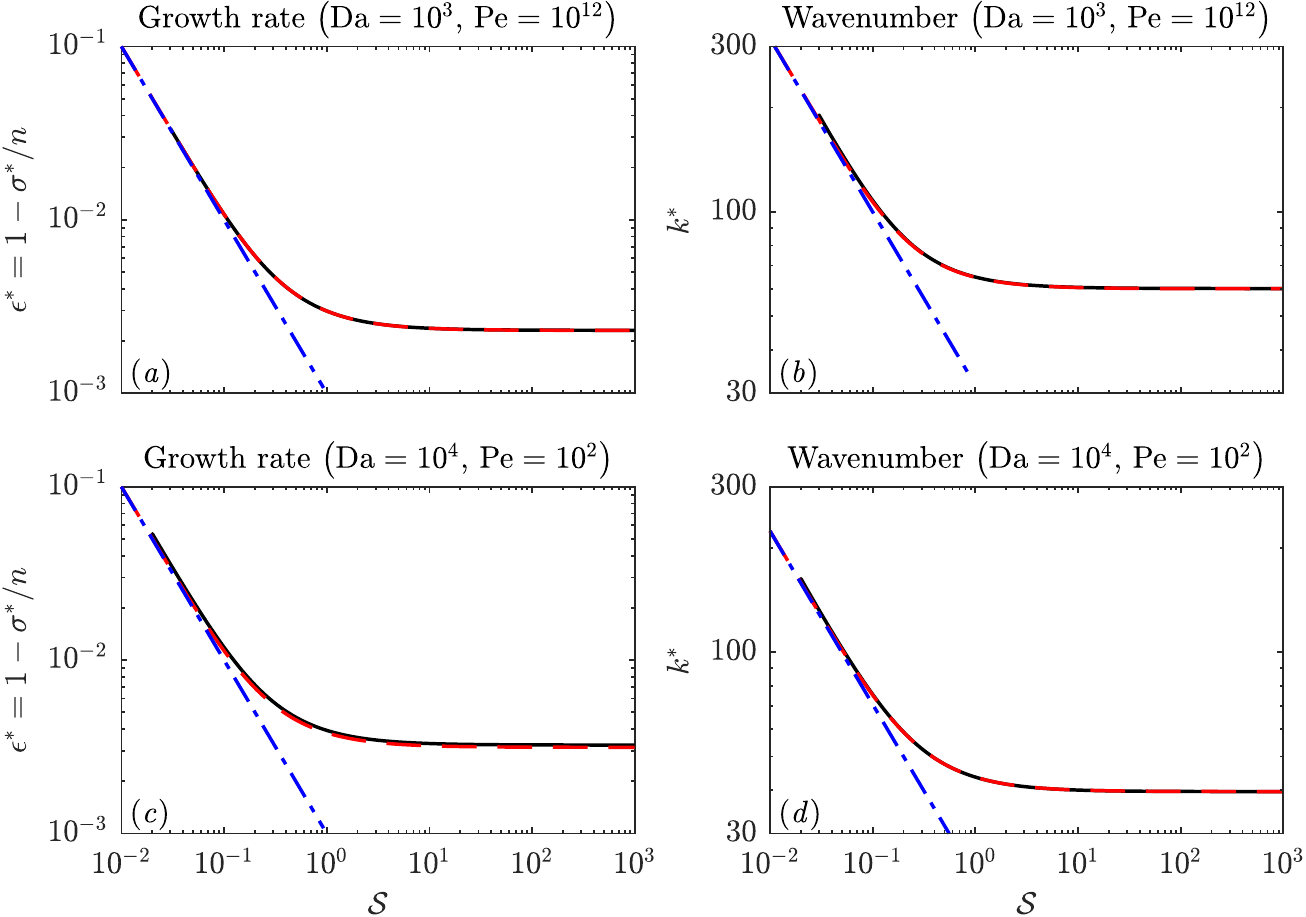}
    \caption{Dependence on matrix compaction (stiffness $\mathcal{S}$) in the two regimes $\mathrm{Pe} \gg \mathrm{Da}$ (panels \textit{a, b}) and $\mathrm{Da} \gg \mathrm{Pe}$ (panels \textit{c, d}). Panels (\textit{a, c}) show the maximum growth rate, and panels (\textit{b, d}) show the corresponding wavenumber. Solid black curves are numerical calculations of the full cubic dispersion relation~\eqref{eq:dispersion3}. Dashed red curves (which are almost indistinguishable) are asymptotic results in the limit of large $\mathrm{Da}$. Blue dot-dashed lines are asymptotic results in the limit of small $\mathcal{S}$. }
    \label{fig:S_scalings}
  \end{center}
\end{figure}

Second we consider the case of diffusion controlled growth ($\mathrm{Da} \gg \mathrm{Pe} $). As before, the growth rate approaches a constant as $\mathcal{S}$ increases, namely
\begin{subequations}
  \label{eq:highDa_highS_limit}
\begin{align}
    \hat{k}^* &\rightarrow (\pi/2)^{1/2} , \\
    \hat{\epsilon}^* &\rightarrow \pi .
  \end{align} 
\end{subequations}
For $\mathcal{S}\ll 1$, as before, $a\sim \mathcal{S}^{-1}$ and we obtain
\begin{subequations}
 
\begin{align}
    \hat{k}^* &\sim (2 \mathcal{S})^{-1/2}, \\
    \hat{\epsilon}^* &\sim\mathcal{S}^{-1}.  \label{eq:highDa_S_scaling}
  \end{align} 
\end{subequations}
These asymptotic results are consistent with numerical results
(figure~\ref{fig:S_scalings}).
Indeed, figure~\ref{fig:S_scalings} shows that compaction reduces the growth rate and increases the wavenumber of the most unstable mode relative to the rigid-medium limit $\mathcal{S} \to \infty$. 
The numerical calculations show that the rigid-medium limit is approximately attained when $\mathcal{S} \gtrsim 1 $. 

The scalings for wavenumber and growth rate are the same in terms of the power-law dependence on $\mathcal{S}$. In either case, a compactible medium is less unstable than a rigid medium. That is, compaction stabilises the system. We can interpret equations \eqref{eq:highDa_infinitePe_S_scaling} and \eqref{eq:highDa_S_scaling} in terms of a critical stiffness such that the instability occurs when ${\mathcal{S}} \geq {\mathcal{S}}_\mathrm{crit}$ where
\begin{subequations}
\begin{align}
    {\mathcal{S}}_\mathrm{crit} &\propto  \frac{1}{\mathrm{Da }} 
                  \qquad (\mathrm{Pe} \gg \mathrm{Da}), \label{eq:Scrit_scaling1} \\
    {\mathcal{S}}_\mathrm{crit} &\propto
            \frac{1}{\sqrt{\mathrm{Da Pe}}}
            \qquad(\mathrm{Da}\gg\mathrm{Pe}). \label{eq:Scrit_scaling2}
  \end{align} 
\end{subequations}
The critical stiffness occurs when the destabilising influence of reaction balances the stabilising influence of compaction
(see \S\ref{sec:physical-discussion-2}).

We can also estimate the aspect ratio $\mathcal{A}$ of the instability for the case $\mathcal{S}\ll1$ by noting that $a \sim \mathcal{S}^{-1}$. The ratio of horizontal to vertical length scale is approximately $\mathcal{A} \sim a/ k^*$. Substituting in the wavenumber scalings, we find
\begin{subequations}
\begin{align}
    \mathcal{A} &\propto  (\mathcal{S} \mathrm{Da })^{-1/2} 
                  \qquad (\mathrm{Pe} \gg \mathrm{Da}), \label{eq:A_scaling1} \\
     \mathcal{A} &\propto (\mathcal{S}^2 \mathrm{Da } \mathrm{Pe })^{-1/4} 
                        \qquad(\mathrm{Da}\gg\mathrm{Pe}). \label{eq:A_scaling2}
  \end{align} 
\end{subequations}
The the horizontal scale of the instability is generally small compared to the vertical scale, but the aspect ratio approaches unity near $ {\mathcal{S}}_\mathrm{crit}$. Thus our assumption that vertical diffusion is negligible compared to horizontal diffusion becomes less valid as we approach $ {\mathcal{S}}_\mathrm{crit}$. However, for the rigid medium case  $ \mathcal{S} \gtrsim 1$, the aspect ratio is always small. 
Furthermore, for the geologically relevant parameters considered in \S\ref{sec:geological_discussion}, the aspect ratio is predicted to be small, \textit{i.e.} the horizontal scale is much smaller than the vertical.

\subsection{Numerical investigation of the critical stiffness} \label{sec:Scrit}
\begin{figure}
  \begin{center}
    \includegraphics[width=1.0\linewidth]{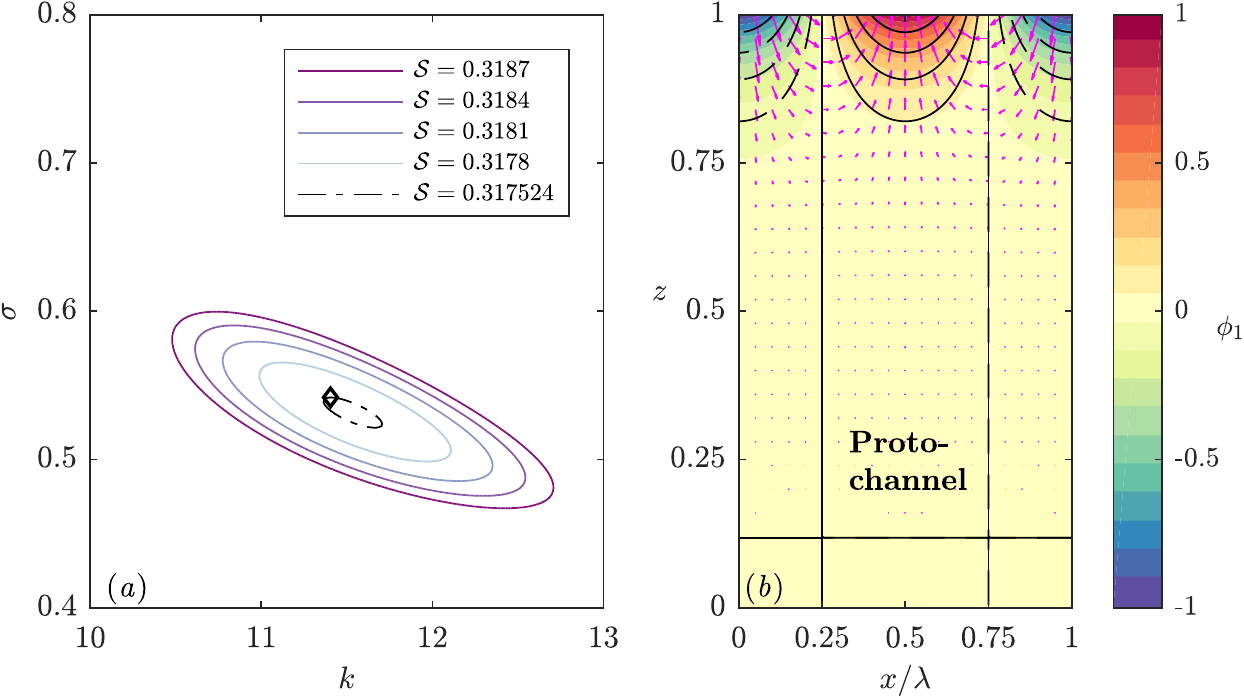}
    \caption{Behaviour as $\mathcal{S} \rightarrow {\mathcal{S}}_\mathrm{crit}^+$ (example with $\mathrm{Da}=\mathrm{Pe}=10$). (\textit{a}) Series of closed loops in $(k,\sigma)$-space as $\mathcal{S}$ decreases toward the critical value [purple to light blue; black dot-dashed loop corresponds to the final iteration; method described in appendix~\ref{app:Scrit}].  
    (\textit{b}) The porosity perturbation $\phi_1$ corresponding to the most unstable mode indicated by the diamond symbol in panel~(\textit{a}). }
    \label{fig:Scrit_explore}
  \end{center}
\end{figure}

We next test these asymyptotic predictions of a critical stiffness by numerically calculating the dispersion relationship at successive values of $\mathcal{S} \rightarrow {\mathcal{S}}_\mathrm{crit}^+$. Figure~\ref{fig:Scrit_explore} shows that (\textit{a})  the dispersion relationship forms closed loops whose size approaches zero; (\textit{b}) the perturbation is localized in an $O(\mathcal{S})$ boundary layer near the upper boundary. The latter observation is consistent with the asymptotic result that the vertical length scale \mbox{$a^{-1} \sim \mathcal{S}$}.  We estimate the critical value $ {\mathcal{S}}_\mathrm{crit}$ using the method described in appendix~\ref{app:Scrit}, and map out the dependence on Damk\"{o}hler number and P\'eclet number.

\begin{figure}
  \begin{center}
    \includegraphics[width=3.4950in]{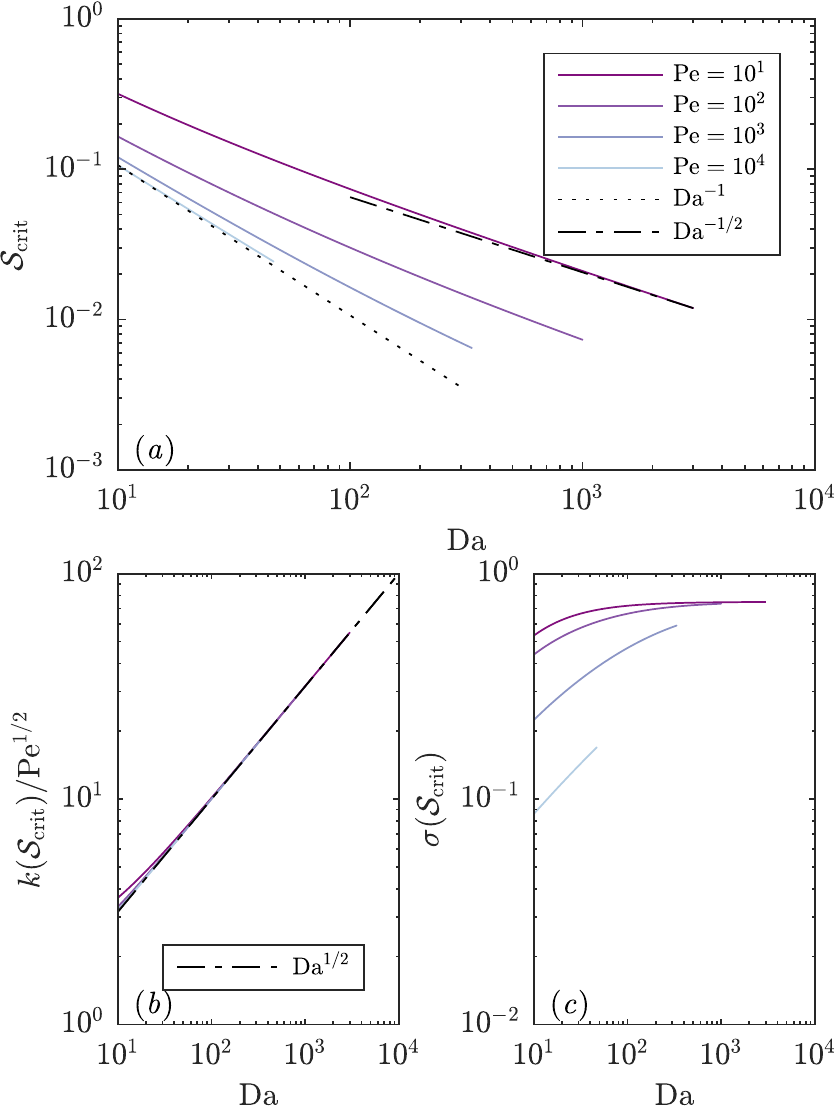}
    \caption{(\textit{a}) The critical stiffness $\mathcal{S}_\mathrm{crit}$ as a function of $\mathrm{Da}$ at $\mathrm{Pe}=10,10^2,10^3,10^4$ (from dark purple to light blue). We also show power law scalings $\mathrm{Da}^{-1}$ (dash-dotted) and $\mathrm{Da}^{-1/2}$ (dotted).  (\textit{b}) The corresponding wavenumber, which obeys the scaling~\eqref{eq:Scrit_k-scaling}. (\textit{c}) The corresponding growth rate, which appears to approach a constant at large $\mathrm{Da}$. }
    \label{fig:Scrit_results}
  \end{center}
\end{figure}

Figure~\ref{fig:Scrit_results}(\textit{a}) shows the dependence of $\mathcal{S}_\mathrm{crit}$ on $\mathrm{Da}$ at $\mathrm{Pe}=10,10^2,10^3,10^4$. The calculations with high $\mathrm{Pe}$ support the prediction of equation~\eqref{eq:Scrit_scaling1} that     ${\mathcal{S}}_\mathrm{crit} \propto  {\mathrm{Da }}^{-1} $ when $\mathrm{Pe} \gg \mathrm{Da}$. The calculations with lower $\mathrm{Pe}$ support the prediction of equation~\eqref{eq:Scrit_scaling2} that     ${\mathcal{S}}_\mathrm{crit} \propto  {\mathrm{Da }}^{-1/2} $ when $\mathrm{Da}\gg\mathrm{Pe}$; they are also consistent with the predicted $\mathrm{Pe}^{-1/2}$ dependence. By estimating the prefactors numerically, we obtain the following scalings:
\begin{subequations}
\begin{align}
    {\mathcal{S}}_\mathrm{crit} &\sim  \frac{1}{\mathrm{Da }} 
                  \qquad (\mathrm{Pe} \gg \mathrm{Da}), \label{eq:Scrit_scaling3} \\
    {\mathcal{S}}_\mathrm{crit} & \sim
            \frac{2}{\sqrt{\mathrm{Da Pe}}}
            \qquad(\mathrm{Da}\gg\mathrm{Pe}). \label{eq:Scrit_scaling4}
  \end{align} 
\end{subequations}
Note that equation~\eqref{eq:Scrit_scaling3} is consistent with the numerical results of \citet{aharonov95} when $\mathcal{S} = O(1)$, although they did not obtain the other limit, equation~\eqref{eq:Scrit_scaling4}.

Figure~\ref{fig:Scrit_results}(\textit{b}) shows that, across the range of parameters considered, the wavenumber at $\mathcal{S}_\mathrm{crit}$ obeys the scaling
 \begin{equation} \label{eq:Scrit_k-scaling}
k(\mathcal{S}_\mathrm{crit}) \sim (\mathrm{Da} \mathrm{Pe})^{1/2}.
\end{equation}
This is the same as the scaling for the large wavenumber cutoff when $\mathrm{Da} \gg \mathrm{Pe}$, identified in \S\ref{sec:wavenumber}. 
However, we see in \S\ref{sec:Scrit_analysis} that a different scaling eventually emerges at higher $\mathrm{Pe}$.

Finally, figure~\ref{fig:Scrit_results}(\textit{c}) shows that the growth rate at $\mathcal{S}_\mathrm{crit}$  appears to approach a (non-zero) constant at large $\mathrm{Da}$, which might be independent of $\mathrm{Pe}$, a hypothesis we confirm in \S\ref{sec:Scrit_analysis}.

\subsection{Analysis of behaviour near the critical stiffness} \label{sec:Scrit_analysis}
We now analyze the structure of the bifurcation at $\mathcal{S}_\mathrm{crit}$. Our goal is to complement the numerical results obtained previously by mapping out the bifurcation structure and obtaining asymptotic results at very large $\mathrm{Da}$ and $\mathrm{Pe}$, regimes that were hard to achieve numerically. 

We proceed by rescaling equations~(\ref{eq:ab_full_1+2}\textit{a},\textit{b}). As we have seen previously, there are two distinguished limits depending on the relative magnitude of $\mathrm{Da}$ to $\mathrm{Pe}$. 

\subsubsection{Case:  $\mathrm{Pe} \gg \mathrm{Da} \gg 1$} 
We first consider the case in which chemical diffusion is negligible. We use the rescaling 
\begin{subequations}
 \begin{align}
      \tilde{x} & =k^2 \mathrm{Da}^{-2}, \\
            \tilde{a} & = a \mathrm{Da}^{-1}, \\
      \tilde{\mathcal{S}} & = \mathcal{S} \mathrm{Da}, \\
 \tilde{\sigma} & = \sigma n^{-1},
  \end{align} 
\end{subequations} 
(extending the scaling first introduced in \S\ref{sec:advection-controlled}).
Then we note that $\mathcal{K} = 1 + k^2/ \mathrm{Da}\mathrm{Pe} = 1+  \tilde{x}(\mathrm{Da}/ \mathrm{Pe})$, so $\mathcal{K}  \sim 1$. We also note that \mbox{$\pi/2 < b<\pi$}, so \mbox{$b^2 \ll a^2 = O(\mathrm{Da}^2)$}. Thus, to leading order, equations~(\ref{eq:ab_full_1+2}\textit{a},\textit{b}) become, respectively, 
\begin{subequations}
\begin{align} \label{eq:ab_ScritA_1}
    \tilde{a} &=  \frac{\left(   \tilde{\mathcal{S}}^{-1} + 
         \tilde{\sigma}       \tilde{x}  \right)}{{2  \left( \tilde{\sigma}
        - \tilde{\mathcal{S}}^{-1}  \right)}}, \\
   \tilde{a}^2 &=\frac{\left( 1- \tilde{\sigma} \right)      \tilde{x}}{{
                \left(  \tilde{\sigma}   - \tilde{\mathcal{S}}^{-1}  \right)}} .  \label{eq:ab_ScritA_2}
  \end{align} 
\end{subequations}
We can eliminate $\tilde{a}$ and rearrange into a quadratic for $\tilde{\sigma}$:
 \begin{equation} \label{eq:quadratic_sigma}
 \tilde{\mathcal{S}}^{2} \tilde{x} (4   + \tilde{x} ) \tilde{\sigma}^2 -2  \tilde{\mathcal{S}} \tilde{x} (1+2\tilde{\mathcal{S}} ) \tilde{\sigma} +(1+4\tilde{x}   \tilde{\mathcal{S}}) = 0.
\end{equation} 
There are repeated roots when the discriminant of the quadratic is zero, corresponding to the left and right hand limits of the loops shown in figure~\ref{fig:Scrit_explore}(\textit{a}). The discriminant is
 \begin{equation}
  \tilde{\Delta} = -16 \tilde{x}   \tilde{\mathcal{S}}^2 \left[ \tilde{x}^2 \tilde{\mathcal{S}} +  \tilde{x} (3 \tilde{\mathcal{S}} -\tilde{\mathcal{S}}^2) +1 \right].
\end{equation} 
Given that $\tilde{\mathcal{S}} >0$ and $\tilde{x}\propto k^2 >0$, the roots $\tilde{\Delta} = 0$ must satisfy the quadratic equation
 \begin{equation} \label{eq:Delta3}
 \tilde{x}^2 \tilde{\mathcal{S}} +  \tilde{x} (3 \tilde{\mathcal{S}} -\tilde{\mathcal{S}}^2) +1 = 0 .
\end{equation} 
The bifurcation (at the critical matrix stiffness) occurs when the discriminant of this quadratic is zero, when 
 \begin{equation}
  0 = (3 \tilde{\mathcal{S}} -\tilde{\mathcal{S}}^2)^2 - 4 \tilde{\mathcal{S}} = \tilde{\mathcal{S}}  (\tilde{\mathcal{S}} -1)^2(\tilde{\mathcal{S}} -4) .
\end{equation} 
The root $ \tilde{\mathcal{S}} = 0$ is excluded because $ \tilde{\mathcal{S}} >0$. The roots $\tilde{\mathcal{S}} = 1$ are excluded because they correspond to repeated roots $\tilde{x} = -1$ in equation~\eqref{eq:Delta3}. The only physically meaningful root is $\tilde{\mathcal{S}} = 4$, which corresponds to repeated roots $\tilde{x} = 1/2$ in equation~\eqref{eq:Delta3}. We substitute back into equation~\eqref{eq:quadratic_sigma} and find that the corresponding $\tilde{\sigma} =1/2$. 

This gives us the critical matrix stiffness and the corresponding properties of the solution (horizontal and vertical wavenumbers and growth rate). In summary, we find that 
 \begin{equation}
  \mathcal{S}_\mathrm{crit}  = 4 \mathrm{Da}^{-1}, \quad
  k_\mathrm{crit}  \sim \mathrm{Da} /\sqrt{2} , \quad
   a_\mathrm{crit}    \sim  \mathrm{Da}, \quad
  \sigma_\mathrm{crit}  \sim  n /2, \quad ( \mathrm{Pe} \gg \mathrm{Da} \gg 1).
  \end{equation} 
In the numerical results (\S\ref{sec:Scrit}), we found the same $\mathcal{S}_\mathrm{crit}  \propto  \mathrm{Da}^{-1}$ scaling, albeit with a different prefactor. However, we didn't observe the $k_\mathrm{crit}  \propto  \mathrm{Da}$ scaling (independent of $\mathrm{Pe}$), which indicates that our numerical calculations were not performed at sufficiently high $\mathrm{Pe}$ to observe the asymptotic regime. Our analysis in this section allows access to that regime. Furthermore, observations such as those in figure~\ref{fig:Scrit_explore} of loops emerging at finite (non-zero) values of the growth rate $\sigma$ emerge as generic features of the bifurcation.  
  
\subsubsection{Case:  $\mathrm{Da} \gg \mathrm{Pe}$}  
We second consider the opposite case in which advection of the 
liquid undersaturation is negligible relative to diffusion. We apply the same type of methodology as before. We use the rescaling 
\begin{subequations}
 \begin{align}
      \hat{x} & =k^2 ( \mathrm{Da} \mathrm{Pe})^{-1}, \\
            \hat{a} & = a ( \mathrm{Da} \mathrm{Pe})^{-1/2}, \\
      \hat{\mathcal{S}} & = \mathcal{S}( \mathrm{Da} \mathrm{Pe})^{1/2}, \\
 \hat{\sigma} & = \sigma n^{-1},
  \end{align} 
\end{subequations} 
(the scaling extends that introduced in \S\ref{sec:diffusion-controlled}).
We note that $\mathcal{K} = 1 + k^2/ \mathrm{Da}\mathrm{Pe} = 1+  \hat{x} $. Again,  \mbox{$b^2 \ll a^2 = O(\mathrm{Da} \mathrm{Pe})$} and $1/\mathrm{Da}\mathcal{S} \sim (\mathrm{Pe}/\mathrm{Da})^{1/2} \ll 1$. Thus, to leading order,  equations~(\ref{eq:ab_full_1+2}\textit{a},\textit{b}) become, respectively, 
\begin{subequations}
\begin{align} \label{eq:ab_ScritB_1}
    \hat{a} &=  \frac{1}{2   \hat{\sigma}  \hat{\mathcal{S}} }, \\
   \hat{a}^2 &=\frac{\left( 1- \hat{\sigma}(1+\hat{x}) \right)      \hat{x}}{
              \hat{\sigma} (1+\hat{x}) } .  \label{eq:ab_ScritB_2}
  \end{align} 
\end{subequations}
We find a quadratic for $\hat{\sigma}$:
 \begin{equation} \label{eq:quadratic_sigmaB}
4 \hat{\mathcal{S}}^{2} \hat{x} (1+ \hat{x} ) \hat{\sigma}^2 - 4  \hat{\mathcal{S}}^2 \hat{x}  \hat{\sigma} +(1+\hat{x}  ) = 0,
\end{equation} 
whose repeated roots are zeros of the discriminant
 \begin{equation}
  \hat{\Delta} = -16 \hat{x}   \hat{\mathcal{S}}^2 \left[ \hat{x}^2  +  \hat{x} (2 -\hat{\mathcal{S}}^2) +1 \right].
\end{equation} 
Again, we have $\hat{\mathcal{S}} >0$ and $\hat{x}\propto k^2 >0$, so roots $\hat{\Delta} = 0$  satisfy 
 \begin{equation} \label{eq:Delta3B}
  \hat{x}^2  +  \hat{x} (2 -\hat{\mathcal{S}}^2) +1.
\end{equation} 
The bifurcation occurs when the discriminant of this quadratic is zero, when 
 \begin{equation}
0 = (2 -\hat{\mathcal{S}}^2)^2 - 4  = \hat{\mathcal{S}}^2  (\hat{\mathcal{S}} +2) (\hat{\mathcal{S}} -2) .
\end{equation} 
The only physically meaningful root is $\hat{\mathcal{S}} = 2$, which corresponds to repeated roots $\hat{x} = 1$ in equation~\eqref{eq:Delta3B}. We substitute back into equation~\eqref{eq:quadratic_sigmaB} and find that the corresponding $\hat{\sigma} =1/4$.  
In conclusion,
 \begin{equation}
  \mathcal{S}_\mathrm{crit}  = 2( \mathrm{Da} \mathrm{Pe})^{-1/2} , \quad
  k_\mathrm{crit}, a_\mathrm{crit} 
  \sim ( \mathrm{Da} \mathrm{Pe})^{1/2} , \quad
  \sigma_\mathrm{crit}  \sim  n /4, \qquad ( \mathrm{Da} \gg \mathrm{Pe}).
\end{equation} 
In the numerical results (\S{}\ref{sec:Scrit}), we found the same scaling relationships (with the same prefactors), so our numerics were able to access this regime adequately. The analysis in this section additionally obtained $\sigma_\mathrm{crit}  \sim  n /4$. 
Therefore, in both regimes we find that the bifurcation at $\mathcal{S}_\mathrm{crit}$ results in an instability with a finite growth rate.

\subsection{Physical discussion of instability mechanism (part II: wavelength selection)} \label{sec:physical-discussion-2}
In \S\ref{sec:physical-discussion-1}, we explained the basic structure of the physical instability mechanism. We found that there is an enhanced vertical flux in proto-channels (regions of positive porosity perturbation) caused by the porosity-dependent permeability. This vertical flux across a background equilibrium concentration gradient dissolves the solid matrix, increasing the porosity, and establishing an instability that grows at a rate $\sigma \sim n$. In this section, we use the insights gained from our asymptotic analysis to explain the physical controls on the vertical and horizontal length scales of the instability, and on the critical matrix stiffness. All of the following estimates are consistent with the results of our asymptotic analysis and numerical calculations. 

We derive scalings focussing on the more interesting case of a compacting porous medium ($\mathcal{S} \ll 1$). 
Results for a rigid porous medium (up to an unknown prefactor) can be obtained by substituting $\mathcal{S} = 1$ into the subsequent scalings, consistent with our numerical results that the rigid medium limit applies when $\mathcal{S} \gtrsim 1$.

First, we consider the vertical length scale of the instability at fixed horizontal wavenumber. 
In a compacting porous medium, mass conservation implies that gradients in porosity are sources or sinks of compaction pressure, as expressed in equation~\eqref{eq:scaling_4}, which we now rewrite by substituting expressions for the base state variables:
 \begin{equation} \label{eq:scaling_5}
 \nabla^2 \mathcal{P}_1 =  
  \frac{n }{\mathcal{S}} \frac{\partial \phi_1}{\partial z}. 
\end{equation} 
We next substitute in the balance between compaction and porosity change from equation~\eqref{eq:solid_1}, namely $\sigma \phi_1 \sim \mathcal{P}_1$, use $\sigma \sim n$, and scale $\partial_z \sim m$, neglecting horizontal derivatives at fixed $k$, 
to obtain
 \begin{equation} \label{eq:scaling_6}
m \sim \mathcal{S}^{-1}.
\end{equation} 
So the vertical structure is controlled by the matrix stiffness.
In the rigid medium case, $m\sim 1$ and the instability extends through the full depth of the melting region.

Second, we consider the horizontal length scale of the most unstable mode $k^*$. We combine equations \eqref{eq:solid_1},~\eqref{eq:scaling_5}~\&~\eqref{eq:scaling_6} to obtain the estimate 
 \begin{equation} \label{eq:scaling_7}
-k^2 \mathcal{P}_1 \sim  \chi_1 / \mathcal{S}^2.
\end{equation} 
Physically, reactive dissolution in the channels requires a convergent flow of liquid into the proto-channels, which must be down a gradient in the compaction pressure.
Then, by substituting \eqref{eq:solid_1} and \eqref{eq:darcy_1} into the liquid concentration equation~\eqref{eq:reaction_1_unmod}, we find that
 \begin{equation} \label{eq:scaling_8}
   \mathcal{S} \frac{\partial \mathcal{P}_1 }{\partial z} - \mathcal{P}_1
   \sim  -  \frac{1  }{\mathrm{Da}} \frac{  \partial \chi_1  }{\partial z }   
     +    \frac{1}{\mathrm{Da}\mathrm{Pe}}\frac{\partial^2  \chi_1 }{\partial x^2}. 
\end{equation} 
Both terms on the left-hand-side are $O(\mathcal{P}_1)$. 
When $\mathrm{Pe} \gg \mathrm{Da}$, diffusion on the right-hand-side is negligible compared to advection of the liquid undersaturation. Then by substituting equation~\eqref{eq:scaling_7} we find
 \begin{equation} \label{eq:scaling_9}
 \quad k^{*2} \sim \frac{\mathrm{Da} }{\mathcal{S}}, \qquad (\mathrm{Pe} \gg \mathrm{Da}).
\end{equation} 
Conversely, if $\mathrm{Da} \gg \mathrm{Pe}$, diffusion dominates the right-hand-side of equation~\eqref{eq:scaling_8} and we find:
 \begin{equation} \label{eq:scaling_10}
k^{*2} \sim \frac{\sqrt{\mathrm{Da} \mathrm{Pe} }}{\mathcal{S}}, \qquad (\mathrm{Da} \gg \mathrm{Pe}).
\end{equation} 
Physically, the perturbed compaction-driven advection against the equilibrium concentration gradient is balanced by either advection or diffusion of liquid undersaturation. 
These results mean that the most unstable horizontal wavelength is proportional to the compaction length (recall that $\mathcal{S} \propto \delta^2$). However, it is much smaller than the compaction length since $\mathrm{Da} \gg 1$, as seen in the 2D numerical calculations of \citet{spiegelman01}.

Third, if the matrix stiffness $\mathcal{S}$ is reduced below some critical value, then the stabilising influence of compaction is dominant over the destabilising influence of reactive melting such that the instability is suppressed. We can obtain an estimate of this critical value as follows.  At the critical value, equation~\eqref{eq:solid_1} gives that compaction balances reaction, so $-\mathcal{P}_1 \sim \chi_1$. 
We next use equation~\eqref{eq:scaling_2}, to obtain $-\mathcal{P}_1 \sim n w_0 \phi_1$.
We substitute into equation~\eqref{eq:scaling_4} to obtain
 \begin{equation}  \label{eq:scaling_11}
{\mathcal{S}}_\mathrm{crit} \sim m/k^{*2},
\end{equation} 
We then substitute in our estimates of the vertical ($m$) and horizontal ($k^{*}$) wavenumbers to obtain:
\begin{subequations}
 \begin{align}
    {\mathcal{S}}_\mathrm{crit} &\sim  \frac{1}{\mathrm{Da }}, 
                  \qquad (\mathrm{Pe} \gg \mathrm{Da}), \\
    {\mathcal{S}}_\mathrm{crit} &\sim
            \frac{1}{\sqrt{\mathrm{Da Pe}}},
            \qquad(\mathrm{Da}\gg\mathrm{Pe}).
  \end{align} 
\end{subequations}
Conversely, we could interpret equation~\eqref{eq:scaling_11} in terms of a minimum wavenumber for growth
 \begin{equation}  \label{eq:scaling_12}
k^2_\mathrm{min} \sim m/\mathcal{S} \sim 1/\mathcal{S}^2 \quad  \Rightarrow \quad k_\mathrm{min}  \sim 1/\mathcal{S}.
\end{equation} 
Thus the maximum wavelength for the instability is proportional to $\beta \delta^2 / \alpha$: the product of the compaction length and the amount of reactive melting over a compaction length. 
In the rigid medium limit ($\mathcal{S} \gg 1$) the vertical wavelength is the full height of the domain ($m\sim 1$). Thus
 \begin{equation}  \label{eq:scaling_13}
k^2_\mathrm{min} \sim m/\mathcal{S} \sim 1/\mathcal{S} \quad  \Rightarrow \quad k_\mathrm{min}  \sim 1/\mathcal{S}^{1/2}.
\end{equation} 

\section{Geological discussion} \label{sec:geological_discussion}
Geologically significant predictions of this model include the conditions under which the reaction-infiltration instability occurs and the size and spacing of the resulting channels. We found earlier that the length scale of the reaction-infiltration instability can be limited by either advection or diffusion. To cover both of these regimes, it is instructive to introduce a reactive length scale $L_\mathrm{eq}$
 \begin{equation}
L_\mathrm{eq} = \left\{
    \begin{array}{ll}
     L_{w} \equiv \frac{\phi_0 w_0}{\alpha R} , 
      & \mathrm{Pe} \gg \mathrm{Da}\text{ (advection controlled)}, \\[2pt]
       L_{D}\equiv 2 \left(\frac{\phi_0 D}{\alpha R}\right)^{1/2},        
      & \mathrm{Da} \gg \mathrm{Pe} \text{ (diffusion controlled)}.
    \end{array} \right.
\end{equation} 
$L_w$ is the distance a chemical component is transported by the background liquid flow over the reaction timescale. $L_D$ is the distance a chemical component can diffuse in the liquid over the reaction timescale.
The factor of 2 is introduced to simplify the dimensional estimates given later in this section.
$L_\mathrm{eq}$ is a generalization of the length scale introduced by \citet{aharonov95}.
The condition for the advection-controlled instability (rather than the diffusion-controlled case) is $ \mathrm{Pe} \gg \mathrm{Da}$. This is equivalent to the statement $L_D \ll L_{w}$, and thus $L_\mathrm{eq} \sim \max \left( L_{w}, L_D\right)$. With this definition, the most unstable wavelength $\lambda^*$ for the instability can be written
 \begin{equation} \label{eq:dim_wavelength}
\lambda^* = \left\{
    \begin{array}{ll}
     2\pi \lambda_c (L_\mathrm{eq} H)^{1/2}, &  \mathcal{S} \gtrsim 1, \\[2pt]
     2\pi \delta (L_\mathrm{eq} \beta / \alpha)^{1/2},  &  \mathcal{S} \ll 1.
    \end{array} \right.
\end{equation} 
In the former equation, we introduced a prefactor $\lambda_c$ that, as $\mathcal{S}\rightarrow \infty$, satisfies $\lambda_c \rightarrow 0.5268$ in the case $\mathrm{Pe} \gg \mathrm{Da}$, and   $\lambda_c \rightarrow \pi^{-1/2} \approx 0.5642$ in the case $\mathrm{Da} \gg \mathrm{Pe}$.
Note that if the reaction rate were infinitely fast (\textit{i.e.} if the liquid chemistry were at equilibrium), then the equilibrium length scale would be zero, and the channels would be arbitrarily small.  
This potentially explains why channels localize to the grid scale in some numerical calculations based on an equilibrium formulations \citep[for example,][]{hewitt10}.

The vertical length scale of the instability $\lambda_v$ is approximately
 \begin{equation} \label{eq:dim_verticallength}
\lambda_v \sim \left\{
    \begin{array}{ll}
     H, &  \mathcal{S} \gtrsim 1, \\[2pt]
     \delta^2 \beta / \alpha,  &  \mathcal{S} \ll 1.
    \end{array} \right.
\end{equation} 
Thus the channels occupy the full depth of the melting region in the case of a rigid medium ($\mathcal{S} \gtrsim 1$) and have a length proportional to the square of the compaction length when $\mathcal{S} \ll 1$. 
The condition $ \mathcal{S} \ll 1$ delineates the compaction-limited instability. In dimensional terms,
  \begin{equation}
  \mathcal{S} \ll 1 \quad  \Leftrightarrow \quad \delta \ll \left( \frac{\alpha H }{\beta} \right)^{1/2}.
\end{equation} 

We also identified the critical condition for the instability to occur. This condition can be written in terms of a critical compaction length $\delta_\text{crit}$ and the reaction length $L_\mathrm{eq}$:
 \begin{equation} \label{eq:critical_compaction_length}
\delta > \delta_\text{crit} \propto  \left( \frac{\alpha L_\mathrm{eq} }{\beta} \right)^{1/2}.
\end{equation} 
Note that \citet{aharonov95} claim that the instability occurs when the compaction length is much larger than the reaction length. Equation~\eqref{eq:critical_compaction_length} shows that the relevant length scale is $\left( {\alpha L_\mathrm{eq} }/{\beta} \right)^{1/2}$, which depends on both the reaction length $L_\mathrm{eq}$ and also on the solubility gradient $\beta$. Indeed, the numerical results of \citet{aharonov95} are consistent with equation~\eqref{eq:critical_compaction_length}.

We now seek to identify the region in parameter space relevant for the partially molten upper mantle. We provide geologically plausible parameter values in table~\ref{tab:parameters}.  Using the central estimates of these parameters, the dimensionless parameters considered previously can be estimated as follows. Damk\"{o}hler and P\'eclet numbers $\mathrm{Da} \approx \mathrm{Pe} \approx 8\times10^7$, reactivity $\mathcal{M} \approx 0.16$, and matrix stiffness $\mathcal{S} \approx 2.5\times10^{-5}$, consistent with our asymptotic approximations $\mathrm{Da},\mathrm{Pe}\gg1$, $\mathcal{M}\ll 1$. 
The aspect ratio of the most unstable mode is $\mathcal{A} \approx 0.02$, consistent with our assumption that the channels are much narrower in the horizontal than in the vertical. 

Returning to dimensional units, if $H = 80$~km (the total depth of the primary melting region; melting may occur deeper in the presence of volatile species) and $\beta = 2 \times 10^{-6}$~m$^{-1}$ \citep{aharonov95} and $\alpha \approx 1$, then  $\left({\alpha H }/ {\beta} \right)^{1/2} \approx 200$~km. The compaction length is typically smaller than this in the mantle, so the instability is likely to be limited by compaction, rather than the total height $H$. For the compaction-limited instability $(\mathcal{S} \ll 1)$, the most unstable wavelength is proportional to the compaction length and the square root of the amount of  chemical disequilibrium that occurs over the height $L_\mathrm{eq}$. The case of the rigid medium is rather different. Here, the most unstable wavelength is the geometric mean of  $L_\mathrm{eq}$ and the total height $H$, and is independent of the solubility gradient $\beta$.

\begin{table}
  \begin{center}
\def~{\hphantom{0}}
  \begin{tabular}{lcll}
      Variable (unit)  & Symbol   &   Estimate (range)   \\[3pt]
      Permeability exponent   & $n$ & $3$  $(2-3)$  \\
      Solubility gradient (m$^{-1}$)  & $\beta$ & $2\times10^{-6}$ ($10^{-6}-4\times10^{-6}$)   \\
      Compositional offset & $\alpha$ & $1$     \\
      Melting region depth (m) & $H$ & $8\times10^{4}$  \\
      Compaction length (m) & $\delta$ & $10^{3}$  ($3\times10^2-10^4$)  \\  
      Melt flux  (ms$^{-1}$)  & $\phi_0 w_0$ & $3\times10^{-11}$ ($5\times10^{-12}-2\times10^{-10}$) \\
      Diffusivity (m$^2$s$^{-1}$)  & $\phi_0 D$ & $3\times10^{-14}$ ($10^{-15}-10^{-12}$) \\
      Reaction rate (s$^{-1}$)  & $R$ & $3\times10^{-8}$ ($10^{-11}-10^{-4}$) \\      
  \end{tabular}
  \caption{Estimates of parameter values with units specified (where relevant), following \citet{aharonov95} as far as possible. For some variables, we consider a range of values to illustrate the range of possible behaviours. This reflects both uncertainty in the parameters themselves, and differences between geological settings. The extreme uncertainty in $R$ reflects uncertainty in the linear chemical dissolution rate and the internal surface area available for reaction. The estimate of $\beta$ is based on thermodynamic calculations \citep{kelemen95b}.}
  \label{tab:parameters}
  \end{center}
\end{table}

Based on the range of parameter values in table~\ref{tab:parameters}, we suggest that $5\times10^{-8} \leq L_{w} \leq 20$~m, and $6\times10^{-6} \leq L_{D} \leq 0.6$~m. The overlap of these ranges suggests that both cases of advection- or diffusion-controlled instability are geologically relevant. In either case, $6\times10^{-6}  \leq L_\mathrm{eq} \leq 20$~m. The corresponding range of critical compaction length is  $2 \leq \delta_\text{crit} \leq 3 \times 10^{3}$~m. If the reaction is fast ($L_{w}$ is small, so $L_\mathrm{eq}$ and $\delta_\text{crit}$ are small), the critical compaction length is likely below the compaction length in the mantle (perhaps 300~m to 10~km), and the instability occurs. However, if the reaction is slow ($L_{w}$ and hence $L_\mathrm{eq}$ and $\delta_\text{crit}$ are large), the critical compaction length may be less than the compaction length, which would suppress the instability. Assuming that the geological observations support channelisation allows us to estimate a lower bound on the reaction rate. We estimate a minimum reaction rate of $R_\mathrm{min}\approx \phi_0 w_0 / \beta \delta^2 \approx 1.5 \times 10^{-11}$~s$^{-1}$ based on the central parameter estimates in table~\ref{tab:parameters} (and a range $R_\mathrm{min}\approx 3\times10^{-14}-2\times10^{-9}$~s$^{-1}$).

We next estimate the dominant wavelength. If the instability does occur, it is most unstable at a wavelength that is smaller than the compaction length by a factor $2\pi (L_\mathrm{eq} \beta / \alpha)^{1/2}$, where $1.5 \times 10^{-5} \leq 2\pi(L_\mathrm{eq} \beta / \alpha)^{1/2}  \leq 5.5 \times 10^{-2}$. Thus the wavelength of the most unstable mode is much smaller than a compaction length. For example, a compaction length of 1~km would have a preferred spacing of 1.5~cm to 55~m. However, the upper end of this estimate corresponding to high $L_\mathrm{eq}$ has a critical compaction length of about 2~km, so the instability would be suppressed. Taking this into account, the largest wavelength expected would be around 25~m. As another example, a compaction length of 10~km would have a preferred spacing of 15~cm to 550~m; the critical compaction length is exceeded throughout the range. The even larger estimates of \citet{aharonov95} are associated with the  limit of a rigid medium $\mathcal{S} \gtrsim 1$, which is probably less geologically relevant.

\begin{figure}
  \begin{center}
    \includegraphics[width=1.0\linewidth]{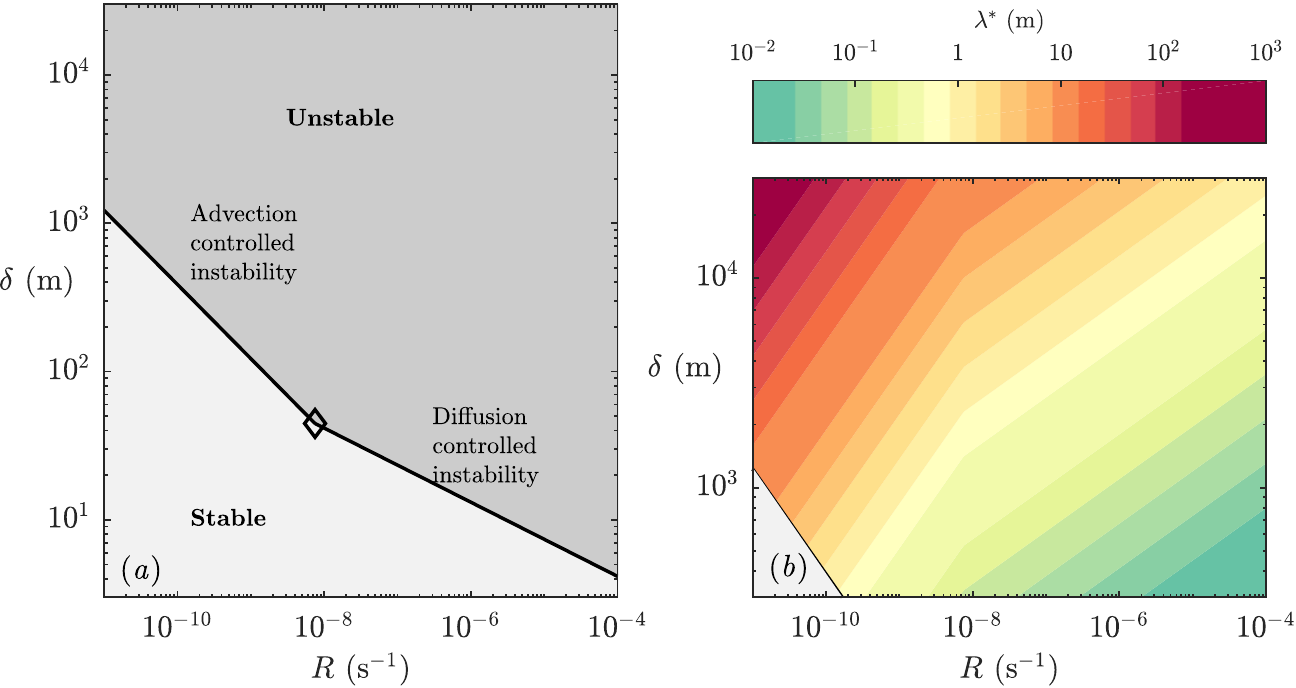}
    \caption{Summary of dimensional predictions. (\textit{a}) The physical compaction length $\delta$ (units m) is generally larger than the critical value $\delta_\text{crit}$ (solid black line), allowing an instability to occur (the darker shaded region). At slower reaction rate $R$ (units s$^{-1}$), the instability may not occur (the lighter shaded region). The diamond marker indicates the transition from advection to diffusion controlled instability, which occurs around $R=10^{-8}$~s$^{-1}$.  (\textit{b}) The most unstable wavelength shown across a physically plausible range of compaction length. The light grey shaded region indicates where the instability is not predicted to occur, as in (\textit{a}).  Unless varied, we use the central estimates of parameter values listed in table~\ref{tab:parameters}. }
    \label{fig:dimensional_predictions}
  \end{center}
\end{figure}

Figure~\ref{fig:dimensional_predictions} summarises the geological implications of our results. 
Figure~\ref{fig:dimensional_predictions}(\textit{a}) shows that the reaction-infiltration instability occurs robustly across a large part of the plausible parameter space (the dark grey region in panel (\textit{a}) covers most of the range of compaction length expected in the upper mantle). The instability is suppressed by small compaction length and slow reaction rate. 
It is also suppressed by a high background melt flux (not shown in figure~\ref{fig:dimensional_predictions}), because the equilibrium length $L_\mathrm{eq}$ increases with melt flux. Figure~\ref{fig:dimensional_predictions}(\textit{b}) shows the predicted horizontal spacing of reactively dissolved channels. Where the instability occurs, we expect it to result in channelized flow on a scale ranging from centimetres to hundreds of meters, a range that is consistent with field observations of reactively dissolved channels \citep{braun02}.

There are additional physical mechanisms, excluded from the present model, that may affect the reaction-infiltration instability. 
First, a greater degree of complexity in the thermodynamic modelling might be important \citep{hewitt10}. For example, volatile chemical species are thought to promote channelized magma flow \citep{keller16} and magma flow can alter the temperature structure \citep{reesjones18a}. 
Second, variation in the background vertical magma flux and solubility gradient \citep{kelemen95b} with depth are very likely to be important, since these drive the instability and control its characteristics. 
Third, rheology also significantly affects the instability. Indeed \citet{hewitt10} used a variable compaction viscosity that suppressed instability, as observed numerically by \citet{spiegelman01}. We discuss this important issue in appendix~\ref{app:hewitt}. Furthermore, it is plausible that reactive channelization is modified by large-scale shear deformation through a viscous feedback \citep{stevenson89, holtzman03a}. 
Fourth, the nonlinear development of the instability and other finite-amplitude effects in the form of chemical and lithological heterogeneity of the mantle may be significant \citep{weatherley12, katz12}. Such heterogeneity may be important because the growth rate of the linear reaction-infiltration instability is relatively slow \citep{spiegelman01}.

\begin{figure}
  \begin{center}
    \includegraphics[width=1.0\linewidth]{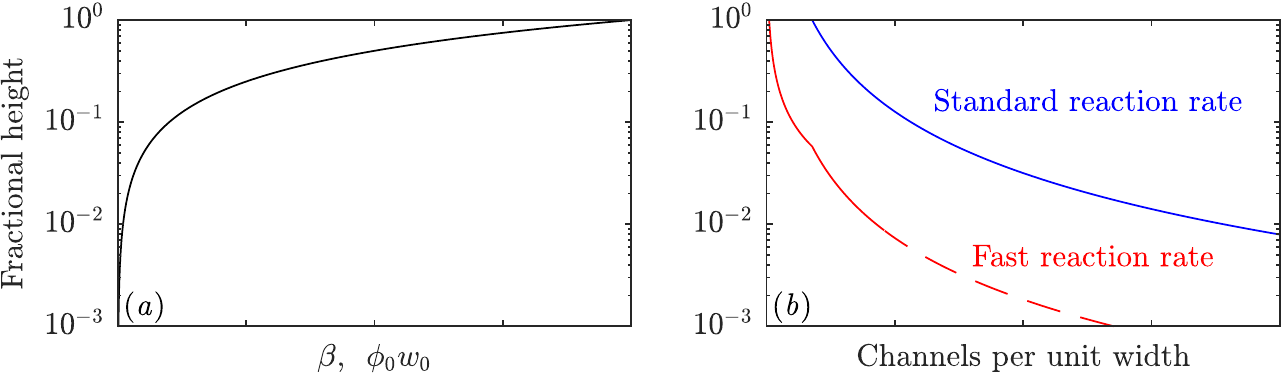}
    \caption{
Speculations regarding the potential consequences of a linear variation in solubility and magma flux with depth, shown in (\textit{a}). 
In (\textit{b}) we show that the number of channels per unit width decreases at shallower depths. 
We use the central estimates of parameter values listed in table~\ref{tab:parameters}, except in the curve marked `Fast reaction rate' ($R = 10^{-10}$~s$^{-1}$). The deepest part of the domain is always stable, although this is only visible in the case of fast reaction rate (plotted as a dashed part of the curve).
This figure is intended to be interpreted qualitatively, so we to do not number the horizontal axis.}
    \label{fig:dimensional_predictions_zdep}
  \end{center}
\end{figure}

We believe that these mechanisms merit detailed study. 
But to speculate about the second of these, we consider a hypothetical situation where the background melt flux and solubility gradient both increase linearly in $z$, as shown in figure~\ref{fig:dimensional_predictions_zdep}(\textit{a}).
Then, our prediction~\eqref{eq:dim_wavelength} gives an estimate of the corresponding most unstable wavelength, shown in figure~\ref{fig:dimensional_predictions_zdep}(\textit{b}).
We find that there are no channels in the deepest part of the domain; channels emerge at shallower depth and progressively coarsen, perhaps due to channel coalescence.
Channel coalescence also occurs in two-dimensional numerical calculations \citep{spiegelman01}, even with a constant solubility gradient, due to the nonlinear development of the instability.
It seems worthwhile to investigate further numerically.

\begin{acknowledgments}
The authors thank M.~Spiegelman, J.~Rudge, D.~Hewitt, I.~Hewitt, A.~Fowler and M.~Hesse for helpful discussions. We thank P.~Kelemen and an anonymous referee for constructive reviews. D.R.J.~acknowledges research funding through the NERC Consortium grant NE/M000427/1. The research of R.F.K.~leading to these results has received funding from the European Research Council under the European Union's Seventh Framework Programme (FP7/2007--2013)/ERC grant agreement number 279925.  We thank the Isaac Newton Institute for Mathematical Sciences for its hospitality during the programme Melt in the Mantle, which was supported by EPSRC Grant Number EP/K032208/1. We also thank the Deep Carbon Observatory of the Sloan Foundation.
\end{acknowledgments}

\appendix
\section{Minimum and maximum wavenumbers} \label{app:kminmax}
The minimum and maximum wavenumbers for instability can be analysed by considering equations~\eqref{eq:critical_condition_2} \& \eqref{eq:ab_full_1+2} which we reproduce here
\begin{align}
0 &= \tan b + b/a, \nonumber \\
    a &=  \frac{\left(     \frac{n\mathcal{K}}{\mathcal{S}} + 
        \frac{\sigma}{\mathrm{Da}} k^2  \right)}{{2  \left( \sigma
        \mathcal{K} - \frac{n}{\mathrm{Da} \mathcal{S}}  \right)}},\nonumber \\
    b^2 + a^2 &=\frac{\left( n-\sigma \mathcal{K} \right) k^2}{{
                \left( \sigma \mathcal{K} - \frac{n}{\mathrm{Da}
                \mathcal{S}} \right)}} . \nonumber
  \end{align} 

Previously we assumed that $\sigma \sim n$, but this assumption can break down near the minimum or maximum wavenumbers. Instead, we make the assumption $ \sigma \mathcal{K}   \gg n /\mathrm{Da} \mathcal{S}$ such that $ n /\mathrm{Da} \mathcal{S}$ can be neglected in the denominators. We verify this assumption \textit{post hoc}. Then
\begin{subequations}
\begin{align} 
    a &=  \left(     \frac{n}{\sigma}\frac{1}{ 2\mathcal{S}} + 
        \frac{k^2}{2\mathrm{Da}  \mathcal{K}}   \right), \\
    b^2 + a^2 &=\left( \frac{n}{\sigma \mathcal{K}} - 1 \right) k^2. 
  \end{align} 
\end{subequations}
We can eliminate $n/\sigma$ between these equations
\begin{equation} \label{eq:general_k}
b^2 + a^2 =\left( \frac{2 a \mathcal{S} }{ \mathcal{K}}  -  \frac{\mathcal{S} k^2}{\mathrm{Da}  \mathcal{K}^2} - 1 \right) k^2. 
\end{equation}

We now simplify these equations for the cases of small and large wavenumber $k$. 
First, when $k$ is small, we assume that $k^2 \ll  \mathrm{Da}  \mathrm{Pe} $  (so $\mathcal{K} \sim 1$)   and $k^2 \ll  \mathrm{Da} / \mathcal{S} $, which again we verify \textit{post hoc}.
Then equation~\eqref{eq:general_k} becomes
\begin{equation} \label{eq:small_k}
b^2 + a^2 \sim \left( {2 a \mathcal{S} } - 1 \right) k^2 \quad  \Rightarrow \quad  k^2 = \frac{b^2 \mathrm{cosec} ^2 b}{ -2  \mathcal{S} b\cot b - 1}.
\end{equation}
Note that $\pi/2< b<\pi$ so $\cot b<0$. 
The minimum wavenumber corresponds to the turning point $d k/db = 0$. 
With some algebra, it is possible to show that this occurs when 
\begin{equation} \label{eq:small_k_bc}
1 - b \cot b + \mathcal{S} b\left[ \cot b + b(1-\cot^2 b) \right] = 0. 
\end{equation}
There is a unique solution $b_c$ to this algebraic equation in  $\pi/2< b<\pi$.

When $\mathcal{S} \gg 1$ (rigid medium), $b_c$ satisfies $ \cot b + b(1-\cot^2 b) = 0$.
We find $b_c \approx 2.2467$, the corresponding $a_c \approx 1.8017$, and
\begin{equation} \label{eq:results_small_k_large_S}
k_\mathrm{min} \sim \frac{1.5171}{ \mathcal{S}^{1/2}}, \quad \sigma(k_\mathrm{min}) \sim 0.2775 \frac{n}{\mathcal{S}}, \quad (\mathcal{S} \gg 1),
\end{equation}
This means that the instability operates at increasingly long wavelength as the matrix rigidity increases. 
Conversely compaction stabilizes the long wavelength limit \citep{aharonov95}.

When $\mathcal{S} \ll 1$ (compactible medium), $b_c$ satisfies $ a \equiv -b\cot(b) = 1/\mathcal{S} $,  and we find 
\begin{equation} \label{eq:results_small_k_small_S}
 \quad k_\mathrm{min} \sim \frac{1}{ \mathcal{S}}, \quad \sigma(k_\mathrm{min}) =  \frac{n}{ 2 }, \quad (\mathcal{S} \ll  1).
\end{equation}    
We substitute all the results back into the assumptions we made and can show that they hold for sufficiently large $\mathrm{Da}$ and $\mathrm{Pe} $. More precisely, when $\mathcal{S} \gg 1$ we need $\mathcal{S} \mathrm{Da}  \mathrm{Pe} \gg 1$ and $\mathrm{Da}  \gg 1$. When  $\mathcal{S} \ll 1$ we need $\mathcal{S}^2 \mathrm{Da}  \mathrm{Pe} \gg 1$ and $\mathcal{S} \mathrm{Da}  \gg 1$.

Finally, we consider the large wavenumber limit. 
Here we assume $k^2 \gg  \mathrm{Da}  \mathrm{Pe} $  (so $\mathcal{K} \sim k^2 /  \mathrm{Da}  \mathrm{Pe}$), $a \gg b$ (since $b < \pi$), and $k^2 \gg S  \mathrm{Da}  \mathrm{Pe}^2$. 
Then  equation~\eqref{eq:general_k} becomes
\begin{equation} \label{eq:large_k}
a^2 =\left( \frac{2 a \mathcal{S}  \mathrm{Da}  \mathrm{Pe}  }{k^2}  - 1 \right) k^2 \quad \Rightarrow \quad k^2 = 2 a \mathcal{S}  \mathrm{Da}  \mathrm{Pe} - a^2.
\end{equation}
The maximum wavenumber corresponds to $d k/da = 0$, i.e. when $a =  \mathcal{S}  \mathrm{Da}  \mathrm{Pe}$, and so 
\begin{equation} \label{eq:results_small_k_small_S}
 \quad k_\mathrm{max} \sim \mathcal{S}  \mathrm{Da}  \mathrm{Pe}, \quad \sigma(k_\mathrm{max}) =  \frac{n}{ 2 } \frac{1}{  \mathcal{S}^2  \mathrm{Da}  \mathrm{Pe}}.
\end{equation}  
This means that compaction stabilizes the system at large wavenumber \citep{aharonov95}. Indeed there is no maximum wavenumber for a rigid medium, instantaneous reaction and/or zero diffusion (infinite  $ \mathcal{S}$, $\mathrm{Da}$ and/or $\mathrm{Pe}$ respectively). However, the growth rate at large $k$ would be infinitessimal.
We check all the assumptions we made and can show that they hold, provided $\mathcal{S} \mathrm{Da} \gg 1$.

\section{Numerical determination of critical matrix stiffness} \label{app:Scrit}
\begin{figure}
  \begin{center}
    \includegraphics[width=0.5\linewidth]{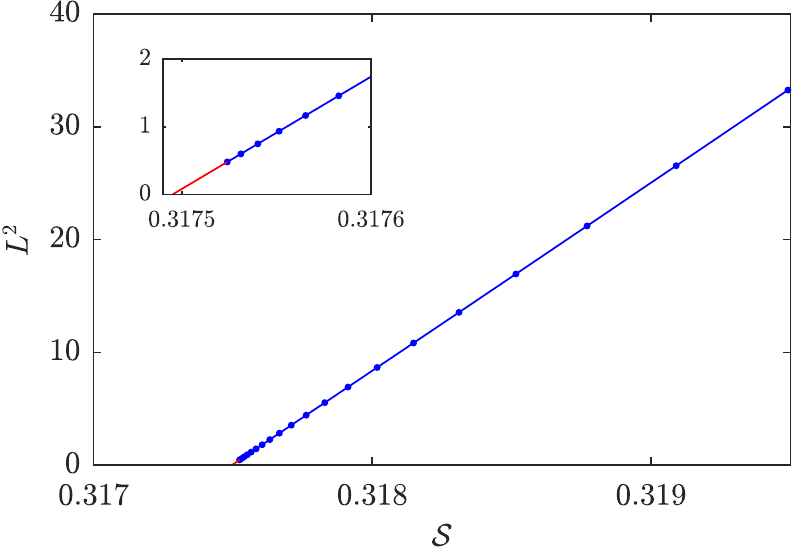}
    \caption{Behaviour as $\mathcal{S} \rightarrow {\mathcal{S}}_\mathrm{crit}^+$ (example with $\mathrm{Da}=\mathrm{Pe}=10$). The length $L$ of the loops shown in figure~\ref{fig:Scrit_explore}(\textit{a}) approaches zero, consistent with equation~\eqref{eq:length_Scrit}. Dots denote values of $\mathcal{S}_n$ according to the iteration~\eqref{eq:Scrit_iteration}. The estimate of  ${\mathcal{S}}_\mathrm{crit}$ is shown using a red line (see inset). }
    \label{fig:Scrit_numerics}
  \end{center}
\end{figure}

When ${\mathcal{S}}$ is sufficiently close to the critical value, the most unstable mode is part of a dispersion curve that forms a closed loop. The size of this loop approaches zero as $\mathcal{S} \rightarrow {\mathcal{S}}_\mathrm{crit}^+$. If we define $L(\mathcal{S})$ as the length of the loop, then we find numerically that 
 \begin{equation} \label{eq:length_Scrit}
L \propto (\mathcal{S} -{\mathcal{S}}_\mathrm{crit})^{1/2}.
\end{equation} 
This behaviour is shown in figures~\ref{fig:Scrit_explore}(\textit{a}) and \ref{fig:Scrit_numerics}.
In \S\ref{sec:Scrit_analysis}, this behaviour emerges as a generic feature of the bifurcation.

Our numerical strategy to determine ${\mathcal{S}}_\mathrm{crit}$ is as follows. Given an initial guess ${\mathcal{S}}_n$, we calculate $L({\mathcal{S}}_n)$. We also estimate $\partial L/\partial{S}$ using a simple finite difference. We then update 
 \begin{equation} \label{eq:Scrit_iteration}
{\mathcal{S}}_{n+1} = {\mathcal{S}}_n - \frac{(1-\lambda) L({\mathcal{S}}_n)}{\partial L/\partial{S}},
\end{equation} 
where $0<\lambda <1$. This is a stabilised Newton iteration, designed to estimate ${\mathcal{S}}_{n+1} $ such that  $L({\mathcal{S}}_{n+1}) \approx \lambda  L({\mathcal{S}}_n)$. The iteration is fastest when $\lambda$ is small but most reliable when $\lambda$ is near 1, so we use $\lambda=0.9$. Motivated by equation~\eqref{eq:length_Scrit}, we next fit a straight line to the square of the length $L^2({\mathcal{S}}_{n})$ (having calculated at least 8 iterates, we use a rolling window of width 8, such that earlier iterates at larger $ \mathcal{S}$ are successively discarded). The intersection of this line with $L=0$ gives an estimate for $\mathcal{S}_\mathrm{crit}$. We iterate until the estimate converges to some small prescribed tolerance ($10^{-8}$). We also calculate the centre of the loop in $(k,\sigma)$-space and extrapolate to $\mathcal{S}_\mathrm{crit}$. Parameter continuation is then used to map out $\mathcal{S}_\mathrm{crit}(\mathrm{Da},\mathrm{Pe})$. This method is robust provided the estimate for $\mathcal{S}$ is sufficiently close to the critical value. This can necessitate taking extremely small steps in parameter space, limiting the calculations that can be performed. 

\section{Technical note on the treatment of reaction rate and compaction viscosity in \citet{hewitt10}} \label{app:hewitt}
\citet{hewitt10} (hereafter H10) argued that the reaction-infiltration instability is not likely to occur in the mantle. This was attributed to a more complex (perhaps more realistic) choice of thermochemical model of melting, leading to a `background' melting rate. However H10 also used a different compaction viscosity compared to our study \cite[and to][]{aharonov95}. In this appendix we argue that the choice of compaction viscosity was largely responsible for the different conclusion, rather than the model of melting. 

The argument made by H10 revolves around the solid mass conservation equation, which (making the same simplifications given in \S\ref{sec:eqs_simplified}, which are also made in H10) can written in dimensionless form as
 \begin{equation}
    \frac{\partial \phi}{\partial t} = \nabla \cdot  \vels  + \Gamma, \label{eq:mass_s_hewitt} \\
\end{equation} 
where $ \nabla \cdot  \vels = \mathcal{P} / \zeta(\phi)$. In the current paper we take $\zeta(\phi)=1$ (non-dimensional version). 
However, H10 takes $\zeta(\phi)=\phi^{-1}$ (non-dimensional version), in which case equation~\eqref{eq:mass_s_hewitt} becomes
 \begin{equation}
    \frac{\partial \phi}{\partial t} = \phi \mathcal{P}  + \Gamma. \label{eq:mass_s_hewitt_2} \\
\end{equation} 
Note that that the compaction pressure $\mathcal{P}$ in our manuscript is equal to the negative of the effective pressure variable in H10. 
Accounting for this sign difference, equation~\eqref{eq:mass_s_hewitt_2} is consistent with equation~(28) in H10. 
Then the growth rate of the linear instability can be estimated 
 \begin{equation} \label{eq:mass_s_hewitt_3}
    \sigma \phi_1 = \phi_0 \mathcal{P}_1 + \phi_1 \mathcal{P}_0   + \Gamma_1.\\
\end{equation} 
H10 argues that the terms $\phi_0 \mathcal{P}_1$ and $\boldsymbol v_{s1} $ (the perturbation to the solid velocity) are small at high wavenumber. 
The thermochemical model of melting used by H10 states that
 \begin{align}
    \Gamma &= \mathcal{G} \left[(1-\phi)\vels + \phi \vell \right]\cdot \hat{\boldsymbol z} 
   \approx \mathcal{G} \left[\vels + \phi^n \hat{\boldsymbol z} \right]\cdot \hat{\boldsymbol z} ,  \label{eq:melt_hewitt} 
\end{align} 
where $ \mathcal{G}$ is a dimensionless melt rate (proportional to our $\beta/\alpha$). Thus perturbations to the melting rate are
 \begin{align}
    \Gamma_1 & = \mathcal{G} \left[\boldsymbol v_{s1}  + n\phi_0^{n-1}\phi_1 \hat{\boldsymbol z} \right]\cdot \hat{\boldsymbol z} 
   \approx n\mathcal{G} \phi_0^{n-1}\phi_1.  
\end{align} 
Equation~\eqref{eq:mass_s_hewitt_3} then becomes
 \begin{equation}
    \sigma \approx   \mathcal{P}_0   + n\mathcal{G} \phi_0^{n-1}, \label{eq:sigma_hewitt} \\
\end{equation} 
which is the same as equation~(32) in H10. 
The steady compaction rate is equal to the steady melting rate:
 \begin{equation}
-\phi_0 \mathcal{P}_0  = \Gamma_0. \label{eq:melt_hewitt_steady} \\
\end{equation}
H10 estimates that the stabilizing compaction term ($\mathcal{P}_0<0$) overcomes the destabilizing reaction term in equation~\eqref{eq:sigma_hewitt}. 
However, it is important to emphasize that the stabilizing term in equation~\eqref{eq:sigma_hewitt} is present only because a strongly porosity-weakening compaction viscosity was chosen. 
A similar effect was also observed numerically by \citet{spiegelman01}. 

What then of the importance of the thermochemical modelling of the reaction rate?
Clearly, a reaction rate parameter appears in equation~\eqref{eq:sigma_hewitt}. 
However, in deriving the approximated melt-rate perturbation $\Gamma_1$ above, H10 shows that perturbations to the liquid flux are dominant over those to the solid flux.
In footnote 3, H10 notes that the previous melting model of \citet{liang10} (which is the same as that of \citet{aharonov95} and hence our own), can be derived from a more general thermochemical model. 
In our notation, this simple melting model has the form
 \begin{equation}
\Gamma =  \phi \vell \cdot \hat{\boldsymbol z} \beta/\alpha.
\end{equation} 
Thus the same form of growth rate estimate as equation~(32) in H10 can be derived using our simplified melting model.
At least for the linear perturbation equations governing the reaction-infiltration instability, the more complex thermochemical model of H10 is not of fundamental importance. 
In this particular context, such a model could be mapped onto our version simply by changing the value of the parameter $\mathcal{G}$.
However, the steady compaction rate given by equation~\eqref{eq:melt_hewitt_steady} does depend on the melting model.

Therefore, with regard to the reaction-infiltration instability, it was the rheology chosen by \citet{hewitt10} that had a decisive effect on the findings, rather than the more complex treatment of melting.

\bibliographystyle{jfm}
\bibliography{main}

\end{document}